\documentclass[fleqn,usenatbib]{mnras}

\usepackage{newtxtext,newtxmath,gensymb,hyperref,caption,subcaption}
\usepackage[T1]{fontenc}

\DeclareRobustCommand{\VAN}[3]{#2}
\let\VANthebibliography\thebibliography
\def\thebibliography{\DeclareRobustCommand{\VAN}[3]{##3}\VANthebibliography}

\usepackage{graphicx}	% Including figure files
\usepackage{amsmath}	% Advanced maths commands
\title[\textit{TESS} RRc stars]{Time series analysis of bright \textit{\textit{TESS}} RRc stars: Additional modes, phase variations and more}

\author[J. M. Benk\H{o} et al.]{
J. M. Benk\H{o},$^{1,2}$\thanks{E-mail: benko.jozsef@csfk.org}
E. Plachy,$^{1,2,3}$
H. Netzel,$^{1,2,4}$
A. B\'odi,$^{1,2}$
L. Moln\'ar$^{1,2,3}$
and A. P\'al$^{1}$
\\
$^{1}$Konkoly Observatory, Research Centre for Astronomy and Earth Sciences, ELKH,
MTA Centre of Excellence,\\
Konkoly Thege Mikl\'os \'ut 15-17, H-1121 Budapest, Hungary\\
$^{2}$MTA CSFK Lend\"ulet Near-Field Cosmology Group\\ 
$^{3}$ELTE E\"otv\"os Lor\'and University, Institute of Physics, P\'azm\'any P\'eter s\'et\'any 1/A, H-1117 Budapest, Hungary\\
$^{4}$ELTE E\"otv\"os Lor\'and University, Gothard Astrophysical Observatory, Szent Imre h. u. 112, 9700, Szombathely, Hungary
}

\date{Accepted 2023 February 16. Received 2023 February 16; in original form 2022 November 29}

\pubyear{2023}

\begin{document}
\label{firstpage}
\pagerange{\pageref{firstpage}--\pageref{lastpage}}
\maketitle

\begin{abstract}
Using two years of data from the \textit{TESS} space telescope, we have investigated the time series of 633 overtone pulsating field RR Lyrae (RRc) stars. The majority of stars (82.8 per cent) contain additional frequencies beyond the main pulsation. In addition to the frequencies previously explained by the $\ell=8$ and $\ell=9$ non-radial modes, we have identified a group of stars where the additional frequencies may belong to the $\ell=10$ non-radial modes. We found that stars with no additional frequencies are more common among stars with shorter periods, while stars with longer periods almost always show additional frequencies. The incidence rate and this period distribution both agree well with the predictions of recent theoretical models. The amplitude and phase of additional frequencies are varying in time. 
The frequencies of different non-radial modes appearing in a given star seem to vary on different timescales.
We have determined a 10.4 per cent incidence rate for the Blazhko effect. For several stars we have detected continuous annual-scale phase change without significant amplitude variation. This type of variation offers a plausible explanation for the ‘phase jump’ phenomenon reported in many RRc stars. The main pulsation frequency could show quasi-periodic phase and amplitude fluctuations. This fluctuation is clearly related to additional frequencies present in the star: stars with two non-radial modes show the strongest fluctuations, while stars with no such modes show no fluctuations at all. The summation of the phase fluctuation over time may explain the O$-$C variations that have long been known for many non-Blazhko RRc stars.
\end{abstract}

%%%%%%%%%%%%%%%%%%%%%%%%%%%%%%%%%%%%%%%%%%%%%%%%%%

%%%%%%%%%%%%%%%%% BODY OF PAPER %%%%%%%%%%%%%%%%%%

\section{Introduction}\label{sec:Intro}

The almost continuous observations of current photometric space telescopes allow us to study 
moderate to small-amplitude brightness variations in stars. 
In the last decade and a half, thanks to the space missions
\textit{MOST} \citep{Walker2003}, 
\textit{CoRoT} \citep{Baglin2006, CorotBook2016}, \textit{Kepler} \citep{Borucki2010} 
and \textit{TESS} \citep{Ricker2015}, 
we have discovered many interesting phenomena,
even in classical radially pulsating variable stars, such as the Cepheid or RR Lyrae  stars.
One of these is the appearance of low-amplitude extra modes in variables pulsating 
in the first radial overtone.

To the best of our knowledge, the first observation of such additional frequencies was made with 
the small Canadian space telescope, \textit{MOST}, in the RRd star (which pulsates in the fundamental and first radial modes simultaneously) AQ Leo \citep{Gruberbauer2007}. 
Shortly afterwards, \citet{Olech2009} found overtone-pulsating (RRc) stars in the
ground-based measurements of the globular cluster $\omega$~Cen whose 
Fourier spectra revealed extra frequencies ${f_x}$, such that the ratio to the dominant frequency ${f_1}$, is 
$f_1/f_x\approx 0.61$. In four RRc stars known at that time in the field-of-view of the \textit{\textit{Kepler}} space telescope, \citet{Moskalik2013, Moskalik2015} presented a complex additional 
frequency structure, and the ratios of the highest amplitude additional frequencies also fell between 0.612 and 0.632.
All RRc and RRd stars observed by \textit{CoRoT} satellite also contain these extra frequencies \citep{Chadid2012,Szabo2014}. 
In parallel with the studies on RR Lyrae stars, similar additional modes have also been found in overtone Cepheids 
in the OGLE survey \citep{Moskalik2009,Soszynski2008,Soszynski2010}. 
In the case of Cepheids, three groups of stars could be identified from 138 Cepheids in SMC,
giving ratios of 0.61, 0.63 and 0.64, respectively. 
These stars are located on parallel sequences in the Petersen diagram \citep{Smolec2016}.
Based on the large OGLE database, \citet{Netzel2015} and \citet{Netzel2019} showed that RRc stars showing extra modes also lie in parallel sequences on the Petersen diagram, similarly to the Cepheids. These sequences are at 0.613, 0.622 and 0.631 period ratios.

The current theoretical explanation of these extra frequencies is that they are associated with non-radial p-modes \citep{Dziembowski2016}. The ratios of overtone Cepheids $f_1/f_x\approx 0.61$, 0.63 and 0.64 belong to $\ell=9, 8$ and 7 modes, respectively. A peculiarity of this explanation is that the observed signals at these ratios are the first harmonics of the true pulsation frequencies. 
It is also true for the frequencies with the ratios 0.61 and 0.63 of RRc stars that they may belong 
to $\ell=9$ and $\ell=8$ non-radial modes.
The sequence at 0.622 of the overtone RR Lyrae stars is explained by linear combination frequencies that appear 
in the spectra. 
(Hereafter, the frequencies with period ratios of $\sim0.61$, $\sim0.62$ and $\sim0.63$ are denoted by 
$f_{61}$, $f_{62}$ and $f_{63}$, respectively.)

\citet{Netzel2015} pointed out that some of the RRc stars also show extra frequencies around the period ratio of 1.458. These frequencies are lower than the fundamental mode frequency of the star. For presentation reasons the inverse ratio ($f_x/f_1\approx 0.686$) is used generally. For these frequencies, no theoretical explanation has yet been found. (These frequencies are hereafter referred to as $f_{68}$.)

The incidence rate of the low-amplitude additional modes in RRc stars
depends strongly on the noise level of the Fourier spectra
which may be reduced not only because of the higher precision of the observations but also 
because of the improved sampling density.
It demonstrates well that
the work that used the largest ground-based observational sample (OGLE survey, \citealt{Netzel2019}) obtained 9.6 per cent for the incidence rate in the Galactic bulge, however, if the higher cadence fields of OGLE were analyzed the incidence
rate proved to be much higher (27 per cents, \citealt{Netzel2015b}).
The space photometric rates, obtained for 
much smaller sample of field RRc stars, are between 81 and 100 per cent \citep{Szabo2014, Moskalik2015, Molnar2015, Molnar2022, Kurtz2016, Sodor2017, Forro2022}

The analysis of space photometric data is not limited to providing a better understanding of the additional modes. 
Continuous space data series can also answer some long-standing questions such as what causes the strong and often 
irregular period (O-C) variations observed in many RRc stars, which cannot be explained by stellar 
evolution (see \citealt{Jurcsik2001, Jurcsik2015} and further references therein).
Or, how the so-called `phase jump' phenomenon happens: the light curves of some RRc obtained in different observing seasons
could not be folded by a common period. However, if we assume that there was a phase jump 
between the observing seasons, the light curves can be folded nicely (e.g. \citealt{Wils2007, Wils2008}).
Is this a real, sudden jump or the result of a continuous phase change 
during the non-observed time interval?
The work of \citet{Moskalik2015} on \textit{Kepler} RRc stars suggests the second scenario, but their sample of four stars is too small for a final answer. Only continuous observation on a larger sample can decide the question.

The observations of the near all-sky
\textit{TESS} mission gives us a good opportunity 
to investigate RRc stars through a large and homogeneous sample of space photometry.

\section{Data and data processing}\label{sec:Data}

\begin{table*}
\centering
\caption{Basic parameters of the used \textit{TESS} sample.
\label{tab:basic}}
\begin{tabular}{lllll}
\hline
Star & RA & DEC & $G_{\rm RP}$ & TIC\,ID \\ 
     & deg & deg & mag & \\
\hline
Gaia DR2 2880528638650410624 &   0.1847089694 & \phantom{$-$}37.8428564459  & 12.73282 &       432552294  \\    
AO Tuc  & 1.0265057115  &  $-$59.4852399803  & 10.849399  &     201252114       \\
2MASS J00120716-5523075 & 3.0299739036 &   $-$55.3854498465 & 13.675732  &     201292299 \\       
V1035 Cas   &    6.9957719417   & \phantom{$-$}49.1625628476  & 11.720211    &   202550541       \\
CO Tuc  & 7.141191811   &  $-$72.16913968   & 13.628787 &      267215550 \\
$\dots$ & & & \\
\hline
\end{tabular}

{The entire table is available in machine-readable format as online supplement.

In Table~\ref{tab:nonRRc} are collected the stars that 
the analysis revealed are not actually RRc stars.
}
%Ez a combined_list_detrend_\textit{TESS}id.txt tablazat.
% A combined_table1.zsh csinalja
% Az lc_nevek.detrend-ben (tenylegesen elemzett *-ok listaja) is 666 kulonbozo *van
\end{table*}

In this paper we used measurements collected by the \textit{TESS} (Transiting Exoplanet Survey Satellite) photometric space telescope.
The detailed specifications of the satellite were presented by \citet{Ricker2015}. Here we just briefly summarize the most important facts.

NASA has launched \textit{TESS} in spring 2018. It uses its four wide-angle CCD cameras to scan 24$\degree$ wide bands of the sky, one $96\degree\times24\degree$ area at a time, known as sectors. 
\textit{TESS} orbits the Earth in a distant and highly elongated orbit (with a perigee of 108\,000 km and an apogee of 375\,000 km, and an orbital eccentricity of 0.55). The orbital period is 13.7~d and one sector is observed in two consecutive orbits. In its first year of operation, the space telescope surveyed 13 sectors of the southern (ecliptic) sky, and in its second year it continued its work in the northern sky.
This work uses data from the first two complete years of observations, i.e., 26 sectors. 
Since sectors have overlapping areas, many of our
targets have data from multiple sectors, some of which are located in the continuous viewing zones near the ecliptic poles that were observed for nearly over a year.

The integrated exposure time of the cameras for so-called full frame images (FFIs) was 30 minutes for the first two years. Selected targets were also observed with a faster (2 min) cadence but in this work only FFI data were used.

\subsection{The sample}\label{sec:sample}

Our primary goal was to detect and analyse the frequency content 
of RRc stars on a large and homogeneous space photometric sample. 
Our selection is based on the catalog of \citet{Clementini2019}, which lists more than 40\,380 RRc stars across the sky, as
observed by the \textit{Gaia} space mission \citep{gaia2016}. Since the additional modes are usually associated with low amplitude frequencies, we restrict ourselves to using the best quality data.
The quality of the \textit{TESS} time series depends primarily on the brightness of the star and the crowding. Therefore, we selected RRc stars brighter than $G_{RP}=$14 mag measured by \textit{TESS} for our sample. 
(Note that the throughput function of \textit{TESS} cameras is close to the function of the $G_{RP}$ broadband filter of \textit{Gaia},
therefore, we use it in this paper as a representation
of apparent brightness.) The brightness of the brightest observed 
RRc star (CS\,Eri, $G_{RP}=8.672$~mag) is well bellow the saturation limit of \textit{TESS} photometry ($\sim7.5$~mag, \citealt{Ricker2015}). These brightness limits
seemed to be a good compromise: the resulting sample is large enough (747 stars), as the noise starts to increase rapidly above 14 mag \citep{Lund2021}.

Besides the observational noise there is another effect that can cause
significant degradation in the data quality, namely the scattered light from the Earth and the Moon. The correction of this effect often needs considerably more effort than a simple detrending, therefore we excluded stars with strong systematics, and focused on the best quality light curves. This selection resulted in a sample of 666 elements.
We then added four additional bright RRc stars which are
missing from the list of \citet{Clementini2019} but have been 
observed by \textit{TESS}. The final 670 stars
of our sample are listed in (the electronic version of) Table~\ref{tab:basic},
sorted by right ascension. The table shows the traditional name (if it
exists, col.~1), celestial position from the
\textit{Gaia} DR2 data base (cols.~2--3),  
the apparent brightness in the \textit{Gaia} $G_{\rm RP}$-band (col.~4) and
for convenience, we include the \textit{TESS} Input Catalog (TIC) IDs \citep{Stassun2019} in the last column.
Since many stars were observed more than one sectors, 1357 time series were analysed in total.

The distribution of our sample in the sky is illustrated 
in Fig.~\ref{fig:Galactic_distr}. 
Stars are missing in areas where \textit{TESS} has not measured 
in its first two seasons: 
along the plane of the ecliptic and a part of the northern ecliptic hemisphere, where excessive scattered light from the Earth or Moon made observations impossible.

\subsection{Data processing}\label{sec:process}

We used the differential-image pipeline developed by \citet{Pal2012MNRAS} to produce the FFI light curves, in which the key element is an image convolution step that is able to correct for many of the instrumental effects. This tool has already been applied to \textit{TESS} RR Lyrae stars in the study of \citet{Molnar2022}. For a more detailed description of the pipeline we refer to \citet{Plachy2021}, where it was used on Cepheid stars. Since our analysis is sensitive to contamination from neighboring stars, we chose a relatively small circular aperture with 1.5 pixel radius even if this led us to an underestimation of the amplitude of the brightness variation. We derived the flux zero-point with respect to the \textit{Gaia} $G_{\rm RP}$ magnitudes \citep{Gaia_rp}. For further corrections we applied a detrending algorithm based on polynomial fitting that was optimized via phase dispersion minimization of the light curves \citep{Bodi2022}, and removed outliers with a sigma-clipping process.
The sigma parameter has been determined for each light curve individually to be 3, 4 or 5 depending on which value gave the best result, based on visual inspection.
For most of the stars we have data from multiple sectors: these light curves were stitched together, with brightness differences between sectors corrected by scaling and shifting.

\section{Fourier analysis of the light curves}

\subsection{The process and its results}

\begin{figure}
\includegraphics[width=0.42\textwidth, angle=270]{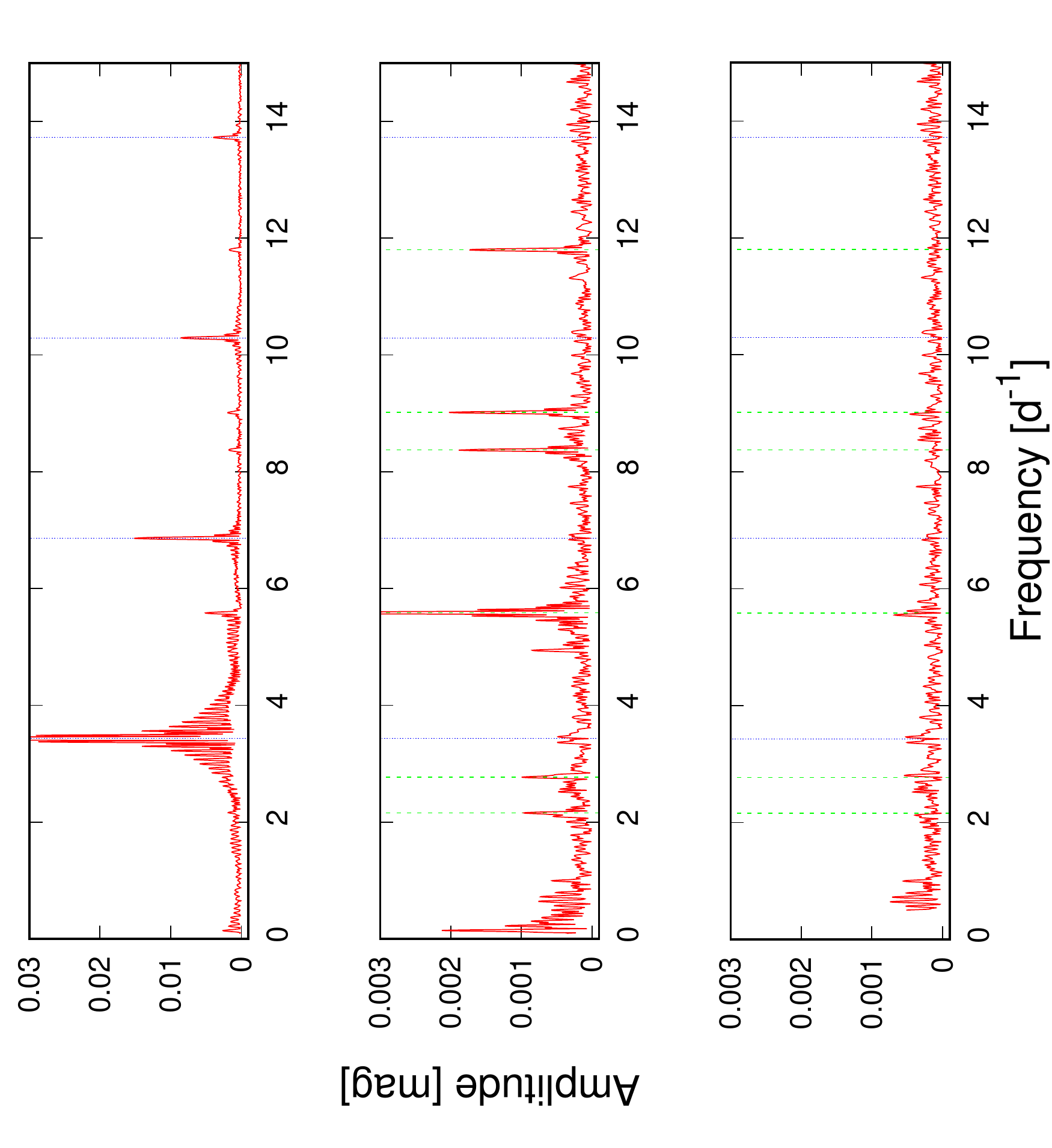}
\caption{Main steps of the Fourier analysis. 
The Fourier spectrum of the light curve of GV~Her from Sector 24 (top).
The blue dotted vertical lines show the location of the main pulsation 
frequency and its harmonics. In the middle is the pre-whitened spectrum 
and the significant additional frequencies found therein 
(indicated by green dashed lines). The residual spectrum (bottom) is
pre-whitened with all frequencies indicated. Note the different amplitude scales.
\label{fig:pre-whitening}}
\end{figure}

Numerous suitable software packages are available for Fourier analysis of space photometric data of RR Lyrae stars. 
Whichever has been tested for this purpose they all gave the same results for frequencies, 
amplitudes and phases within error \citep{Chadid2010, Benko2010}. 
Due to its simple automation, the {\sc SigSpec} package 
\citep{SigSpec, SigSpec_Users_Manual} was chosen for the present analysis.
We ran {\sc SigSpec} twice on each data set. First, we determined the dominant pulsation frequency $f_1$ by fitting of its significant harmonics (see Table~\ref{tab:Four_res}), then we pre-whitened the data series with 
the dominant frequency and these harmonics, and searched consecutively for all significant frequencies ($S_f >5$) in the residual, where $S_f$ means the spectral significance defined in \citet{SigSpec}. In the first run, the frequency search was performed between 0 and the Nyquist frequency (24~d$^{-1}$). 
As the Fourier analysis was performed for the 27-day sectors, in several cases even the 
4th or 5th harmonic was not significant (especially for sinusoidal light curves). 
In the majority of cases, we fitted five harmonics, but sometimes we had to reduce the 
number to four or even three to keep the fit stable.
Since the highest amplitude additional frequencies tend to locate around $f_1$, 
and the computing time is strongly dependent on the length of the frequency interval,
in the second run, both the low- and high-frequency parts were truncated between 0.5 and 15~d$^{-1}$.
Fig.~\ref{fig:pre-whitening} illustrates the process with one of 
the data sets for GV\,Her while numerical 
results are shown in Table~\ref{tab:Four_res} and Table~\ref{tab:Fr_add}.

\begin{table*}
\centering
\caption{Results of the Fourier analysis}\label{tab:Four_res}
\begin{tabular}{cllllllll}
\hline
Star & Sec. & $f_1$ & $A_1$ & $R_{21}$ & $R_{31}$ & $\phi_{21}$ & $\phi_{31}$ & Remark \\
 &  & d$^{-1}$ & mag & & & rad & rad &  \\
\hline
2MASS J00120716-5523075  & 3 & 3.19373 & 0.0971 & 0.216 & 0.052 &  3.042 & $-$0.394  & Bl,$f_{68}$,$f_{61}$,blend? \\
2MASS J02253626-5519227  & 2 & 2.58304 & 0.1037 & 0.091 & 0.071 &  3.798 & \phantom{$-$}1.025   & $f_{68}$,$f_{61}$,$f_{63}$ \\
2MASS J03132028-1726447  & 1 & 3.80346 & 0.0974 & 0.166 & 0.036 &  3.035 & $-$0.267 & - \\
2MASS J03354606-2902102  & 1 & 3.60580 & 0.0999 & 0.215 & 0.067 &  3.025 & $-$0.487  & $f_{61}$ \\
2MASS J04274606-6202513  & 12 & 3.39956 & 0.1161 & 0.199 & 0.050 & 3.014 & $-$0.439  & - \\
$\dots$ & & & & & & & & \\
\hline
\end{tabular}

The entire table is available in machine-readable format as online supplement.
%Ez a table_2.tex tablazat (table2.zsh gyartja le.) 670 csillag.
% Bl es egyebek bevitele: table2_Bl.zsh --> table_2_new.tex
\end{table*}

\begin{figure*}
\includegraphics[width=0.3\textwidth, angle=270]{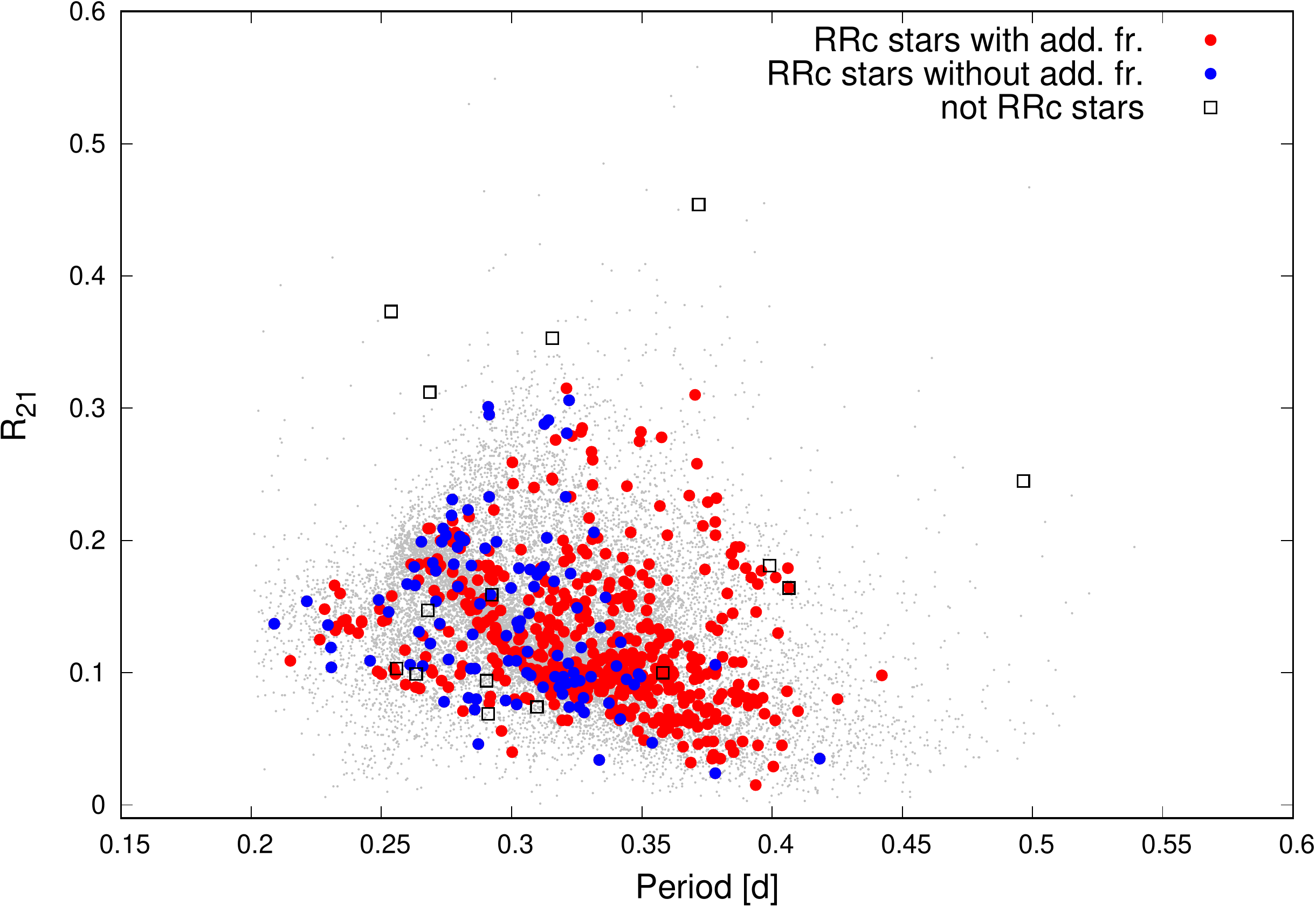}
\includegraphics[width=0.3\textwidth, angle=270]{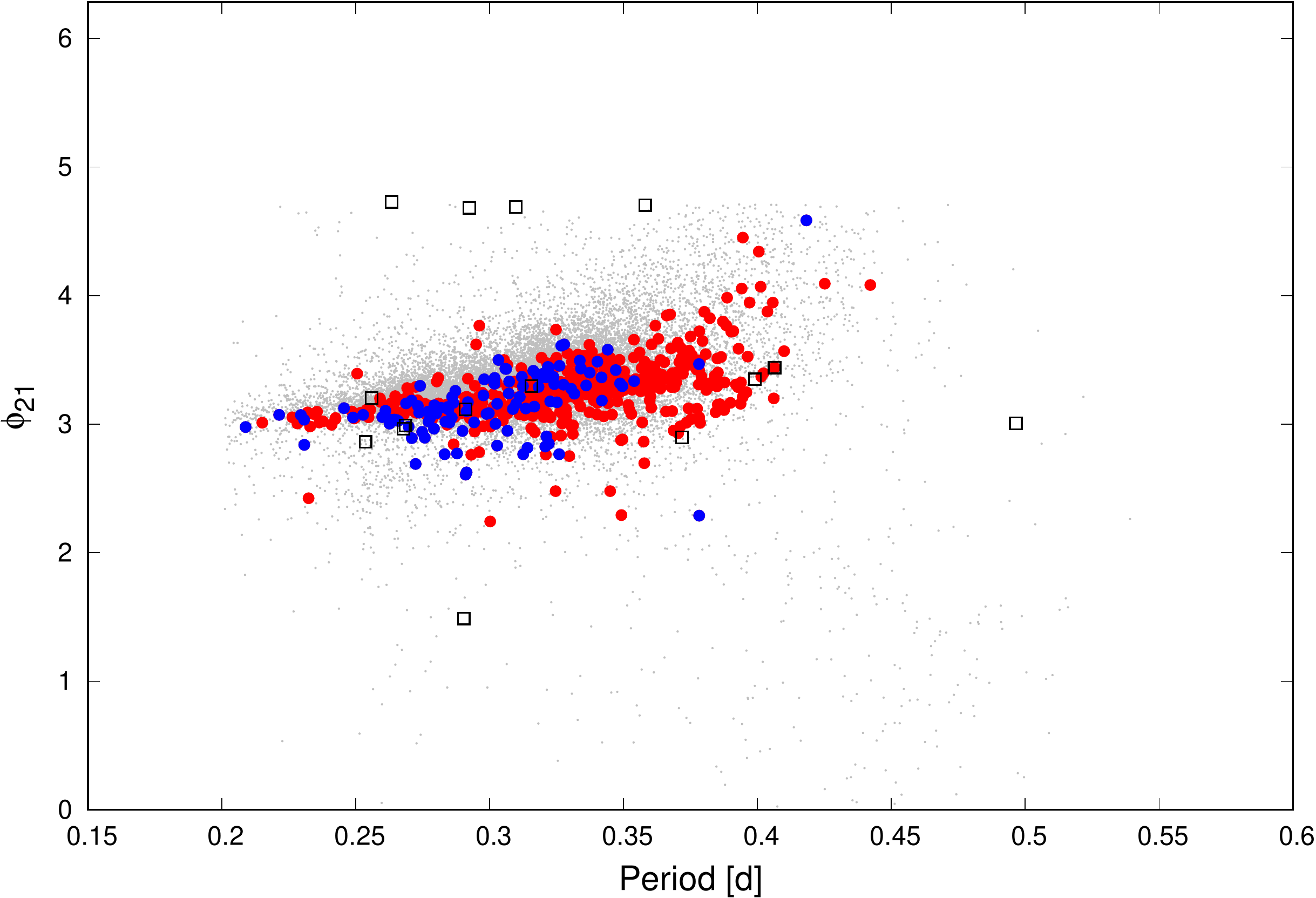}

\includegraphics[width=0.3\textwidth, angle=270]{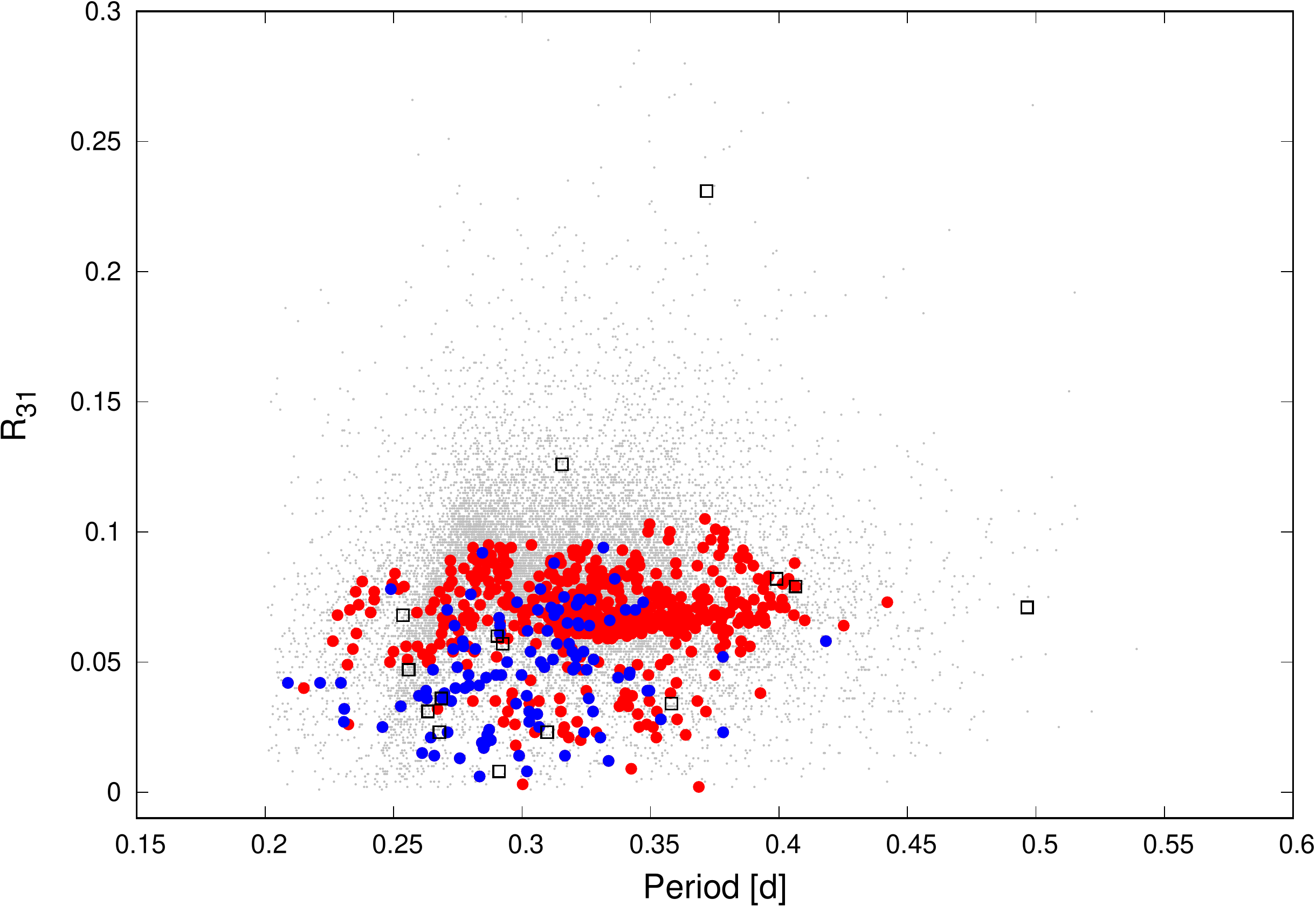}
\includegraphics[width=0.3\textwidth, angle=270]{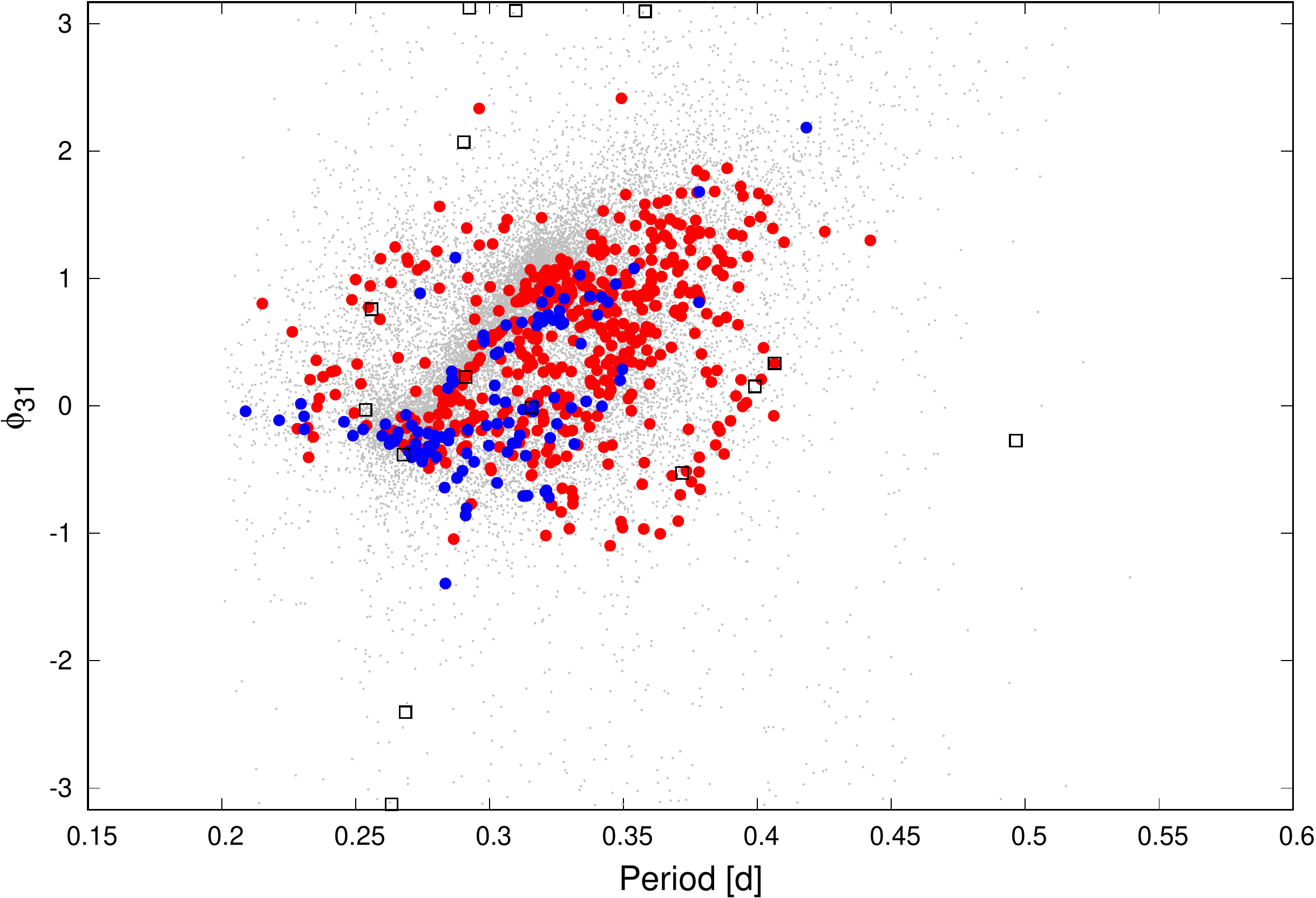}
\caption{Fourier parameters $R_{21}$, $R_{31}$, $\phi_{21}$ and $\phi_{31}$ as a 
function of the pulsation period $P_1$.
The red and blue symbols indicate stars that do and do not show additional frequencies, respectively. The empty black rectangles show the identified non-RRc stars listed in Table~\ref{tab:nonRRc}. 
For comparison, OGLE RRc stars are also shown (gray dots).}
\label{fig:Fourier_param}
\end{figure*}

Table~\ref{tab:Four_res} contains the names of the stars (col.~1), the
number of sectors used (col.~2), the dominant pulsation frequency (col.~3) and its Fourier amplitude ($A_1$, col.~4).
The Fourier parameters in cols.~5--8
are defined in the usual way: $R_{21}=A_2/A_1$, $R_{31}=A_3/A_1$, 
$\phi_{21}=\phi_2-2\phi_1$ and $\phi_{31}=\phi_3-3\phi_1$ \citep{Simon1981}. 
Here the amplitudes and phases ($A_i$, $\phi_i$, $i=1, 2, 3$) are derived from sine-based Fourier solutions as
\begin{equation}\label{eq:Fourier}
    m(t-t_0)=m_0+\sum_{i=1}^{N}A_i\sin\left[2\pi if_1(t-t_0)+\phi_i\right], 
\end{equation}
where $m(t)$ is the light curve, $m_0$ the zero point (the average brightness),
$N=3, 4$ or 5. The starting epoch $t_0$ is always the observed time of the first element of the given data sector.
If a star was detected in more than one sector, the frequencies, amplitudes and Fourier parameters 
are calculated from the merged data sets.
The last column contains some remarks. The meaning of these notations is explained 
in Sec.\ref{sec:Bl} and Sec.\ref{sec:Add}.

Table~\ref{tab:Fr_add} shows the results of the Fourier analysis from the
residual light curves after we subtracted a non-linear fit containing the
main frequency and its harmonics. The consecutive pre-whitening and 
frequency search was carried out on each data set separately.
The names of the data sets (col.~1) consist of the name of the star and the number of the \textit{TESS} sector. 
The next columns show the dominant
frequency $f_1$ (col.~2) its Fourier amplitude and phase ($A_1$, $\phi_1$ in cols.~3--4). Additional columns list the frequencies (col.~5), amplitudes (col.~6) and phases (col.~7) associated with the extra peaks detected in the pre-whitening steps, with decreasing amplitudes.
Columns 8 and 9 give the period and amplitude ratios of the given frequencies and the main frequency, respectively.
The last column shows the possible identification of the frequency (see later).

The numerical values in Tables~\ref{tab:Four_res} and \ref{tab:Fr_add}
give the significant decimal digits plus one additional digit.
The frequency accuracy can be first estimated by the Rayleigh frequency resolution value. In our case, 
it is between 0.037 and 0.0027 d$^{-1}$, depending on the length of the data series. 
Practical experience and e.g., the work of \citet{Kallinger2008} show that the Rayleigh resolution strongly overestimates the error of the frequencies. 
On the other hand, the covariance matrix of the non-linear fit specifies only the formal error, 
and thus tends to underestimate the actual error.
Here, we determined the Fourier parameters of the stars separately for each sector, and also 
for the full-length data series, and then assigned the errors obtained by comparing the values from
different sectors to the values obtained from the complete data series. 
So, for example, for frequency $f_1$ we obtained
$\sigma(f_1)$ between $1.0\cdot10^{-6}$ and $5\cdot10^{-4}$~d$^{-1}$. 
That is, the frequency can be determined to an accuracy of about four digits.
We assume that this is also true for the frequencies for stars measured in only one sector. Accordingly, the frequency is given to 5 decimal digits in Table~\ref{tab:Four_res}. 
For the other Fourier parameters we proceed in a similar way.

\subsection{Fourier parameter vs.\ period planes}

In Fig.~\ref{fig:Fourier_param} we plot the $R_{21}$, $R_{31}$, $\phi_{21}$ and $\phi_{31}$ Fourier parameters given in Table~\ref{tab:Four_res} as a function of the dominant period ($P_1=1/f_1$). 

\begin{table*}
\centering
\caption{Additional frequencies in \textit{TESS} RRc Fourier spectra}\label{tab:Fr_add}
\begin{tabular}{crrrrrrrl}
\hline
Data set & $f_1$ & $A_1$ & $f_x$ & $A_x$ & $\phi_x$ & $f_1/f_x$ & $A_x/A_1$ & ident. \\
 & d$^{-1}$ & mag & d$^{-1}$ & mag & rad &  &  & \\
\hline
2MASS\_J00120716-5523075\_s1 & 3.19360 & 0.0963 &  3.09213 & 0.0190 &  0.6752 & 0.968 & 0.198 & $f_1-f_B$\\
2MASS\_J00120716-5523075\_s1 &  $\dots$ & $\dots$ & 3.30145 & 0.0033 &  2.7865 & 0.967 & 0.035 & $f_1+f_B$\\
2MASS\_J00120716-5523075\_s1 & $\dots$ & $\dots$ &  2.16806 & 0.0032 &  1.8787  & 0.679 & 0.034 & $f_{68}$\\
$\dots$ & & & & & & & &  \\
2MASS\_J00120716-5523075\_s2 & 3.19385 & 0.1066 & 3.09200 & 0.0214 & 1.1599 & 0.968 & 0.201 & $f_1-f_B$\\
2MASS\_J00120716-5523075\_s2 & $\dots$ & $\dots$ & 2.17098 & 0.0037 & $-$1.8961 & 0.680 & 0.035 & $f_{68}$\\
2MASS\_J00120716-5523075\_s2 & $\dots$ & $\dots$ & 6.28551 & 0.0036 & 2.8806 & 0.508 & 0.034 & $2f_1-f_B$\\
 $\dots$ & & & & & & & & \\
\hline
\end{tabular}

The entire table is available in machine-readable format as online supplement. 
%Ez az fr_add.table.ident additional_2.0.zsh gyartja le, 
% illetve aztan az sp_ident.zsh azonositja a frekvenciakat.
% A biztosan nem RRc csillagok, ill. a blendek kiszedve.
\end{table*}

\begin{table*}
\centering
\caption{List of identified non-RRc stars}\label{tab:nonRRc}
\begin{tabular}{llrrlc}
\hline
Gaia DR2 ID & RA & DEC & $G_{\rm RP}$ & Star & Note \\ 
& deg & deg & mag &  & \\
\hline
5073663940716363904 &       \phantom{0}46.7627438301 &   $-$26.0423842601 &  9.703495 & HD 19479 & W UMa \\
3295420894600845952 &    \phantom{0}75.0391559681 & \phantom{$-$}12.7591150306 & 13.9129\phantom{00} & & RRab \\
5256789350372915328   & 149.687401418 & $-$59.4632677075  & 12.035355 &  & RRd \\
5202509386185816704  &  149.902329174 & $-$78.7385930078 &  13.073698 &  & Ecl? \\
1066922279223207040    & 152.8130218992 &  \phantom{$-$}67.6083869718 &   13.832476 &  & HADS? \\
5234537300921901824    &171.255477458  &  $-$69.2574940474  &   12.416878 &  & RRab \\
3513784423567794688 &  186.2895988837 & $-$21.6645591793 & 11.9342\phantom{00} & XX Crv & RRd \\
067166601592685440 &  201.7892016557 & $-$53.9859257004 & 13.659541 & BT Cen & RRab \\
 1588420094522305024 & 229.3719083681 &  \phantom{$-$}48.1918167858 &  13.550349 & CRTS J151729.2+481131 & RRd \\
 1342291874024443008 &  264.657398107 &  \phantom{$-$}37.0837733568  & 13.847074 &    CRTS J173837.7+370502 & W UMa \\
2092889154772199168 &   284.9514704246 & \phantom{$-$}36.1800009151 &  12.624741 & & W UMa \\
2089300658056315776 &   302.1361662576 & \phantom{$-$}52.9588098225 &   12.884088 & & W UMa \\ 
2005340708908381696  & 331.6717909237 &  \phantom{$-$}53.8447046694 &   13.663897 &  & W UMa \\
1956531880221332480 & 332.5323465599 & \phantom{$-$}40.9195805444 & 9.764321 & DE Lac & HADS? \\ 
\hline
\end{tabular}
%Ez a combined_list_detrend.txt tablazattal azonos szerkezetu tablazat
\end{table*}

It can be clearly seen that the vast majority of stars lie within a well-defined area in each panel of the figure. 
These areas overlap strongly with the areas defined by 
the OGLE sample \citep{Soszynski2014, Soszynski2016AcA, Soszynski2019},
but some significant shifts can also be found between them. As already noted by
\citet{Molnar2022} the sensitivity function of the \textit{TESS} cameras and the $I$-band filter used by the OGLE team are close but not identical. They found that the $R$ parameters are nearly identical, but there is a phase shift in the $\phi$ parameters. From our much larger element sample, however, a small amount of shift ($\approx0.015$) is also seen for the $R_{31}$ parameter.  

The stars that are located far from the main areas have been examined one by one to identify the causes of their deviant position. 
A common reason is that a nearby star is blended with the RR Lyrae variable in the \textit{TESS} images.
Such stars were verified by examining their unusual light curves, Fourier spectra and raw CCD images. 
For the 22 such identified stars, the word `blend' 
is added to the note column of Table~\ref{tab:Four_res}. These stars are not plotted in Fig.~\ref{fig:Fourier_param} and their frequencies are also not listed in Table~\ref{tab:Fr_add}.
For a further 28 stars, only one of the measured sectors 
suffered from blending or the presence of blending was not clear. 
These stars were marked as `blend?' in Table~\ref{tab:Four_res} and their data were used for the analysis. 
We note that the Fourier parameters of these stars are not significantly different from those of the un-blended stars.

The situation, however, is different for targets that turned out not to be RRc stars. This was the second most common reason for the unusual Fourier parameters. The 14 identified non-RRc stars are collected 
in Table~\ref{tab:nonRRc} where we also give their possible classifications. Their position are shown by empty rectangles in Fig.~\ref{fig:Fourier_param}.
Omitting the non-RRc stars and those that were certainly blended, we ended up with a sample consisting of 633 elements for further analysis.

\section{Analysis of the sample}\label{sec:anal}

\begin{figure}
\includegraphics[width=0.47\textwidth]{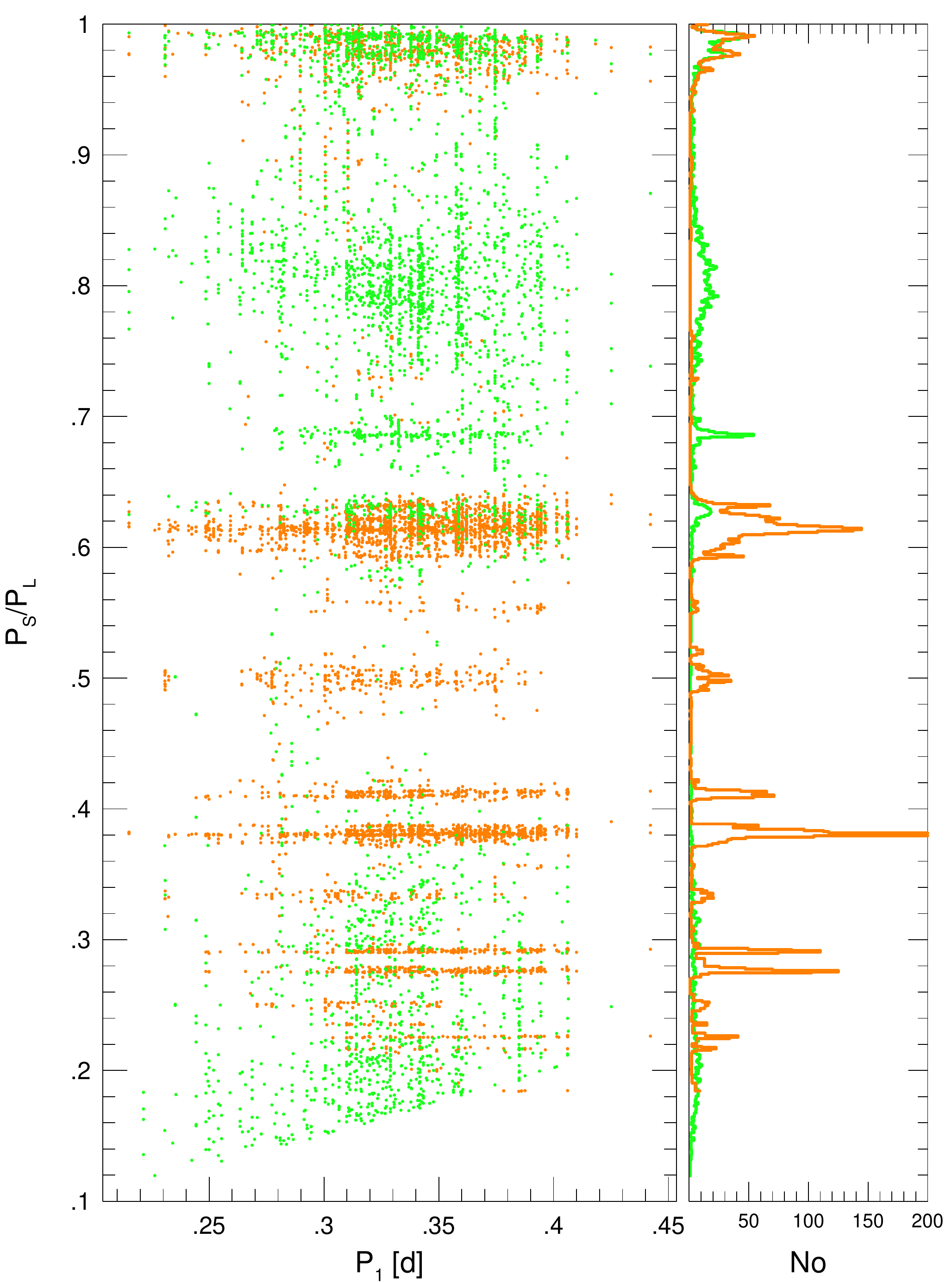}
\caption{Left panel: period ratios of the 
additional mode $P_x$ and the dominant pulsation period $P_1$ as a function of $P_1$.
In accordance with previous studies, we always divide the shorter period $P_{\rm S}$  
with the longer one $P_{\rm L}$, so the values are always between 0 and 1.
Orange symbols: $P_x/P_1<1$ (shorter periods), green symbols: $P_x/P_1>1$ (longer periods).
Right panel: histograms show the additional 
frequency incidence rates as a function of the period ratio.
\label{fig:Petersen}}
\end{figure}
% fr_add.table.ident-bol es az fr_add.rovid.1+, 1- 
%-bol legyartja: fr_add.mgo

Beyond the main pulsation frequency (and its harmonics)
we detected more than 11\,000 significant frequencies in the 633 analysed stars. 
%For 615 of all stars, we found at least one more frequency.

All the detected frequencies are plotted in a Petersen-type diagram in Fig.~\ref{fig:Petersen}.
Each point in the figure belongs to a frequency of a certain 
observed sector of a given star. That is, the frequency content of 
each sector of each star is included here.
We show this because the amplitude of frequencies 
could vary in time and therefore the frequency content of different sectors might be different.
Following earlier studies the period ratios are
calculated as $P_{\rm S}/P_{\rm L}$, where 
$P_{\rm S}$ and $P_{\rm L}$ are the shorter and the longer periods, respectively. 
This representation results in a compact
diagram but it also has disadvantages. In the figure, 
orange symbols indicate $P_x/P_1<1$, while green symbols indicate $P_x/P_1>1$.
%It can be seen that 
Clearly, with large number of frequencies present, the ratios of the distant frequencies fall
close to each other, mixing the distributions and making the interpretation difficult.

The detected frequencies are not randomly distributed in Fig.~\ref{fig:Petersen}, 
but group at certain period ratios.
A histogram of the frequency distribution was constructed by dividing 
both intervals ($P_1>P_x$ and $P_1<P_x$) into 0.001-wide bins and plotting the 
number of frequencies in each bin in the right panel of Fig.~\ref{fig:Petersen}.
In the unstructured part of the figure, each bin of the histogram 
contains 1-3 frequencies.  The level of significant deviation from this random distribution 
was chosen to be three times of this value (9 frequencies per bin). 
Although this choice is somewhat arbitrary, it is a good representation 
of what we see when we look at the figure.

The frequencies were first identified automatically by assuming 
that the spectra of stars may contain similar frequencies 
as seen for \textit{Kepler} stars \citep{Moskalik2015, Sodor2017}.
Namely, (i) Blazhko side-peaks
($f_1\pm f_B$, $2f_1\pm f_B$, etc.), (ii) frequencies of probable
non-radial modes, various harmonics and sub-harmonics of them 
($f_{61}, f_{63}$, $0.5f_{61}$, $1.5f_{61}$, $\dots$), (iii) 
and linear combinations of these frequencies to each other and with the main pulsation frequency.

The automatic frequency identifier works
by taking all the frequency combinations in the
\textit{Kepler} RRc stars listed in the work of \citet{Moskalik2015} 
and calculating their possible values for a given star. 
We then form the difference between the actual peaks $f$
to be identified and the calculated frequencies ($\Delta_i=\vert f - f_i\vert$).  
The differences are sorted and if the smallest
difference $\Delta_i^{\mathrm{min}}$ is less than $0.05$~d$^{-1}$ then we accept that the 
observed frequency is the same as the hypothesized combination.  
This $\Delta_i^{\mathrm{min}}$ value is an estimated sum of the Rayleigh 
frequency resolution ($\sim0.037$~d$^{-1}$)
and the possible difference between the frequencies associated with the actual and the assumed 
average  periodicity ratios (i.e. 0.613). 
The actual value of $\Delta_i^{\mathrm{min}}$ is, however, not crucial, since
the automatic identification was manually checked one by one and
was modified if it was incorrect. 
The result of this process is given in column 10 of Table~\ref{tab:Fr_add} 
as a possible identification. 
A side result of the manual check is that few additional 
non-RRc stars are filtered out. These are also included in 
Table~\ref{tab:nonRRc} (and already excluded in the final 633-element sample).

\subsection{The Blazhko effect}\label{sec:Bl}

Stars showing the Blazhko effect (i.e., correlated amplitude and frequency modulation of the main pulsation)  
are usually searched for in larger surveys by detecting side peaks in 
the Fourier spectra next to the main frequency and harmonics. 
If we search our sample using this criterion, we find side peaks in 250 stars.
These frequencies are located in the dense regions
around $P_{\rm S}/P_{\rm L}\sim 1.0, 0.5, 0.33, 0.25$
in Fig.~\ref{fig:Petersen}.
These ratios correspond to the ratios of the side peaks to 
the main frequency: $f_1/(f_1\pm f_B)$,
$f_1/(2f_1\pm f_B)$,  $f_1/(3f_1\pm f_B)$ and $f_1/(4f_1\pm f_B)$, respectively.

On closer inspection of the light curves and Fourier spectra 
of the stars showing the above frequencies, they can be divided into two 
main groups: (i) In one group the light curves show 
a clear variation in amplitude, and -- more importantly -- if there have been 
observations of the star in several sectors, all their spectra show the same side peak frequencies. 
The side peaks generally form a regular multiplet structure
(triplet, quintuplet, etc) although sometimes side peaks are significant only on one side. 
(ii) In the other group there is no visible amplitude variation, and the distance of the 
side peaks from the main frequency varies from sector to sector, 
and what is more, in some sectors there are no significant side peaks to detect at all. 
It is also common that these spectra do show not regular multiplets but a complex forest of peaks instead.

The first group shares all the characteristics of stars 
with the `classical' Blazhko effect. In total, 68 such 
RRc stars were found in our sample, representing 
10.7 per cent of all stars. These stars are marked with a sign `Bl' in the last 
column of Table~\ref{tab:Four_res}.
The work of \citet{Netzel2018} has recently reviewed the incidence rate of the Blazhko 
effect for different groups of stars (see Table 4 in their paper).
The value for the Galactic field RRc stars is not included in the table, as no study has yet been done 
to analyse the large number of field RRc stars in a homogeneous way.
The missing value can now be added to the table: our rate of 10.7 per cent here is in good agreement 
with the 13 per cent value of \citet{Molnar2022} who used data 
from the first two sectors of \textit{TESS}, only.

The previously obtained values fell 
between 4 per cent (for the LMC)
and 19 per cent (for the globular cluster NGC\,6362). As opposed to the SMC and LMC,  
in the case when all the stars are in the Galaxy, 
all previous studies have shown that the Blazhko 
effect is 3--5 times more frequent in RRab stars than in 
RRc stars.
This remains true even if only space photometric 
measurements are taken into account.  
The incidence rate of Blazhko RRab stars in the 
much smaller \textit{Kepler} sample is 51--55 per cent \citep{Benko2019}, 
while \citet{Molnar2022} obtained a ratio of 47.5--70.7 per cent 
for a subset of \textit{TESS} RRab stars (82 stars), 
\citet{Plachy2019} found 44.7 per cent of 371 K2 RRab stars to be Blazhko stars,
but \citet{Kovacs2018} even found 91 per cent of 151 K2 RRab stars to be modulated.

What about the second group?
As mentioned in the introduction,
RRc stars often contain non-radial modes
(a detailed discussion of these is given in next section). 
The amplitude and value of the frequencies associated with these modes is known to vary
in time \citep{Moskalik2015}.
The quasi-periodic nature of this variation explains 
the forest of side-peaks and the appearance of different `modulation' frequencies 
from sector to sector.
In short, in this case there is no Blazhko effect, 
but we see a reflection of the time variations of the non-radial modes in the Fourier spectra.
The phenomenon is investigated further in Sec.~\ref{sec:CVZ}.

\subsection{Additional frequencies}\label{sec:Add}

In this section, we refer to all significant frequencies that 
do not belong to the main pulsation or the above discussed 
phenomena (the Blazhko effect and low amplitude 
quasi-periodic modulations) as additional frequencies. 
If we omit the identified Blazhko frequencies and frequencies belonging to variations
longer than 10 day periods   
(e.g. instrumental trends, satellite orbit), 524 stars 
(82.8 per cent of the complete RRc sample) remain where we could find frequencies corresponding 
to a signal with a significant amplitude. 
It is emphasized that we were not looking for frequencies with 
a given frequency ratio, but for all significant frequencies 
in the interval between 0.5 and 15~d$^{-1}$.

The incidence rate of the additional mode frequencies found in previous works are quite different from each other.
The rate of the $f_{61}$ additional frequency has been found to be 27 and 63 per cents in the high-cadence OGLE 
fields and in the globular cluster NGC 6362, respectively \citep{Netzel2015b, Smolec2017}.
\citet{Jurcsik2015} obtained a 38 per cent rate for $f_{61}$ stars in the globular cluster M3.
For frequencies with a period ratio between 0.60 and 0.64, the incidence rate in the Galactic 
bulge was 8.3 per cent from the OGLE data, while it was 1.3  per cent for frequencies with a period ratio $\sim0.68$ \citep{Netzel2019}.

Additional mode frequencies have been detected in all previously published (small number of) 
overtone RR Lyrae stars measured by space photometry with \textit{CoRoT} 
and \textit{\textit{Kepler}} \citep{Szabo2014, Moskalik2015, Molnar2015, Kurtz2016, Sodor2017}. 
Each of a small, 4-element sub-sample from the K2 mission also shows additional frequencies \citep{Molnar2015}. 
By using 31 selected stars measured in the first two observing sectors of \textit{TESS} satellite, \citet{Molnar2022} determined the incidence rate of RRc stars with additional modes to be 81 per cent.
The latter value is in good agreement with the 82.8 per cent we obtained. 
That is, it seems that additional frequency mode is present in most RRc stars, but not in all of them.
This is exactly the situation that \citet{Netzel2022} obtained from their theoretical models.

The lower incidence rates of ground-based surveys can be explained by the interaction of several causes.
For example, different detection limits are associated with different brightness of samples measured with detectors of different sensitivities, and the difference in data series length and sampling, and the varying amplitudes of the additional frequencies are also not negligible effects.

\subsubsection{Frequencies connected to $f_{61}$ and $f_{63}$}

\begin{table}
\centering
\caption{Positions and probable identifications of the horizontal ridges in Fig.~\ref{fig:Petersen}}
\label{tab:fr_track}
\begin{tabular}{llll}
\hline
Ratio & Size & No. & Ident. \\
$P_x/P_1$ ($P_1/P_x$)      & $\times$0.001 &  &  \\
\hline
0.216	&	3	&	56	&	$3f_1+f_{61,62,63}$ \\ %
0.226	&	3	&	94	&	$2f_1+1.5f_{61}$ \\ %
0.235	&	1	&	30	&	$f_1+2f_{61}$\\ %
%0.250	&	6	&	89	&	4f_1\pm f_B \\
0.272	&	1	&	20	&	$2f_1+f_{60}$; $3f_1+f_{68}$ \\ %
0.275	&	6	&	377	&	$2f_1+f_{60,61,62,63}$ \\ %
0.291	&	5	&	329	&	$f_1+1.5f_{61}$; $f_1+f_{61}+0.5f_{63}$ \\ %
0.297	&	2	&	22	&	$f_1+1.5f_{63}$ \\ %
%0.307	&	1	&	8	&	2f_{61} \\ %
%0.329	&	1	&	19	& 	3f_1 + f_B \\
%0.332	&	2	&	38	&	3f_1\pm f_B \\
%0.336	&	4	&	68	&	3f_1\pm f_B \\
%0.356	&	1	&	6	&	2f_1+0.5f_{61} \\
0.372	&	4	&	42	&	$2f_1+f_{68}$ \\ %
0.376   &   5   &   125 &   $f_1+f_{60}$ \\ %
0.380	&	7	&	662	&	$f_1+f_{61}$ \\ % 
0.383	&	3	&	312	&	$f_1+f_{62}$ \\ %
0.387	&	5	&	146	&	$f_1+f_{63}$ \\ %
0.410	&	3	&	217	&	$1.5f_{61}$ \\ %
0.412	&	4	&	153	&	$f_{61}+0.5f_{63}$ \\ %
0.417	&	2	&	27	&	$1.5f_{63}$; $f_1+2f_{68}$ \\ %
%0.491	&	3	&	40	& 	2f_1\pm f_B\\
%0.497	&	6	&	124	&	\\
%0.502	&	6	&	122	&	\\
%0.509	&	2	&	23	&	\\
%0.520	&	2	&	23	&	\\
%0.536	&	1	&	12	& 	? \\
0.552	&	3	&	26	&	$f_1+0.5f_{61}$ \\ %
0.557	&	3	&	19	&	$f_1+0.5f_{63}$ \\ %
0.593	&	10	&	149	&	$f_1+f_{68}$ \\ %
0.602	&	20	&	311	&	$f_{60}$ \\ %
0.613	&	18	&	990	&   $f_{61}$ \\ %
0.622	&	9	&	534	&	$f_{62}$ \\ %
0.631	&	12	&	341	&	$f_{63}$ \\ %
0.729	&	3	&	17	&	$2f_{68}$ \\ %
%0.958	&	3	&	23	&	? Bl-ben van, szimmetrikus!\\
%0.965	&	4	&	64	&	? \\
%0.976	&	10	&	270	&	1.5f_{68}? \\
%1.000	&	9	&	457	&	f_1\pm f_B \\
%1.024	&	0	&	1	&	? \\
1.197 (0.835) &   33  &   208 & $0.5f_{60}$ \\ %
1.229 (0.814) &   54	&	329	& $0.5f_{61}$ \\ %
1.264 (0.791) &   30  &   420 & $0.5f_{63}$ \\ %
1.459 (0.685)	&	10	&	321	& $f_{68}$ \\ % 
1.594 (0.626)	&	37	&	226	& $f_{61}-f_1$ \\ %
\\
\hline
\end{tabular}
\end{table}

\begin{figure}
\includegraphics[width=0.25\textwidth, angle=270]{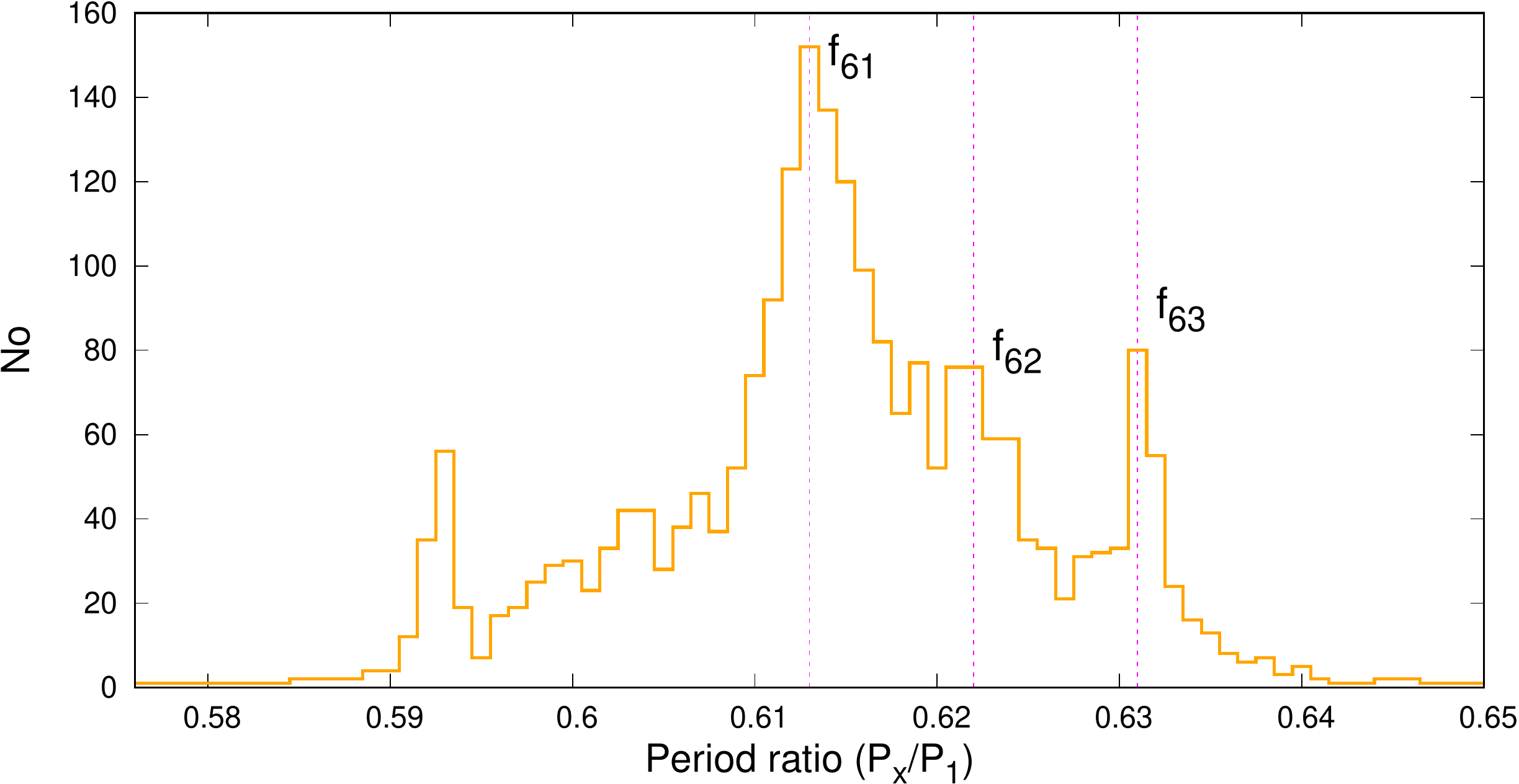}
\caption{
Distribution of the frequencies around the period ratio $P_x/P_1=0.61$. The dashed vertical lines show the positions the three ridges found in OGLE bulge data \citep{Netzel2019}.}\label{fig:zoom_hist}
\end{figure}
In Table~\ref{tab:fr_track} we listed the most frequent period ratios 
viz.\ the visible horizontal ridges in
Fig.~\ref{fig:Petersen}.
The columns of the table show the period ratios $P_x/P_1$ and $P_1/P_x$ in brackets where 
$P_x>P_1$ (col.~1); the size (thickness) 
of each ridge expressed in 0.001-wide bins (see Sec.~\ref{sec:anal}), (col.~2); 
the number of points counted in the bins of col.~2, i.e., 
the total number of frequencies assigned to a given ratio, (col.~3); 
and the identification of the frequencies of the ridge (col.~4).

The most populated horizontal group of frequencies is at the period ratio  
of $P_x/P_1=0.613$ (denoted by $f_{61}$ in Table~\ref{tab:Fr_add} and \ref{tab:fr_track}). 
We detected this frequency
in 360 stars (57 per cent of the sample).
As mentioned in the introduction, this additional frequency was the first to be identified in the 
Fourier spectrum of stars pulsating in the radial overtone mode. According to the
theoretical model of \citep{Dziembowski2016}, this frequency is 
caused by an $\ell=9$ non-radial mode. 
To be precise, this frequency is the
harmonic of the actual mode frequency. 
That is, the actual mode 
is associated with the frequency of $0.5f_{61}$. 
This frequency was also observed in 186 stars. 
In many cases both the $f_{61}$ and $0.5f_{61}$ components are significant, but sometimes only one of them is.
The only reason for this phenomenon is that
we look at stars from different viewing angles, 
as shown by the work of \citet{Netzel2021}.
The third harmonic ($1.5f_{61}$)
is detectable in 125 stars, and the fourth harmonic ($2f_{61}$) 
is also significant in the spectra of 8 stars.
It is known that the $0.5f_{61}$ frequencies do not appear at exactly 
half of $f_{61}$, which is caused by the time dependence of the frequencies 
\citep{Szabo2010, Moskalik2015}. 
It differs from star to star, and this is probably 
the reason that the ridges of half frequencies are 
on average twice as wide as those of the $f_{6x}$ frequencies 
(see col~2. in Table~\ref{tab:fr_track}).

In addition to the harmonics of the mode frequency, we also identified
a number of linear combinations with the main 
pulsation frequency, $f_1$.
Probable identifications of the combination frequencies 
corresponding to the period ratios of the horizontal ridges 
shown in Fig.~\ref{fig:Petersen} is given in the last column 
of Table~\ref{tab:fr_track}. 
In some cases, these ridges are formed by several 
frequencies of nearly equal ratios.

A zoom of the histogram in Fig.~\ref{fig:Petersen}
around the ratio of 0.61 is shown in Fig~\ref{fig:zoom_hist}. 
The distribution shows a composite structure: it is neither symmetrical nor simple, 
but has two sharp `side-peaks' as well, 
one at $P_x/P_1=0.631$ and the other at 0.593.
The frequencies forming the 0.631 ratio (hereafter $f_{63}$) 
are also known from previous works \citep{Moskalik2015, Netzel2019}.
The explanation is also similar to that of $f_{61}$: 
$0.5f_{63}$ is the 
frequency of the non-radial mode $\ell=8$ and $f_{63}$ 
is its harmonic. The mode frequency itself ($0.5f_{63}$), and 
several linear combination frequencies are also common in our sample, resulting in clearly visible horizontal ridges in Fig.~\ref{fig:Petersen}.

\begin{figure}
\includegraphics[width=0.45\textwidth, angle=0]{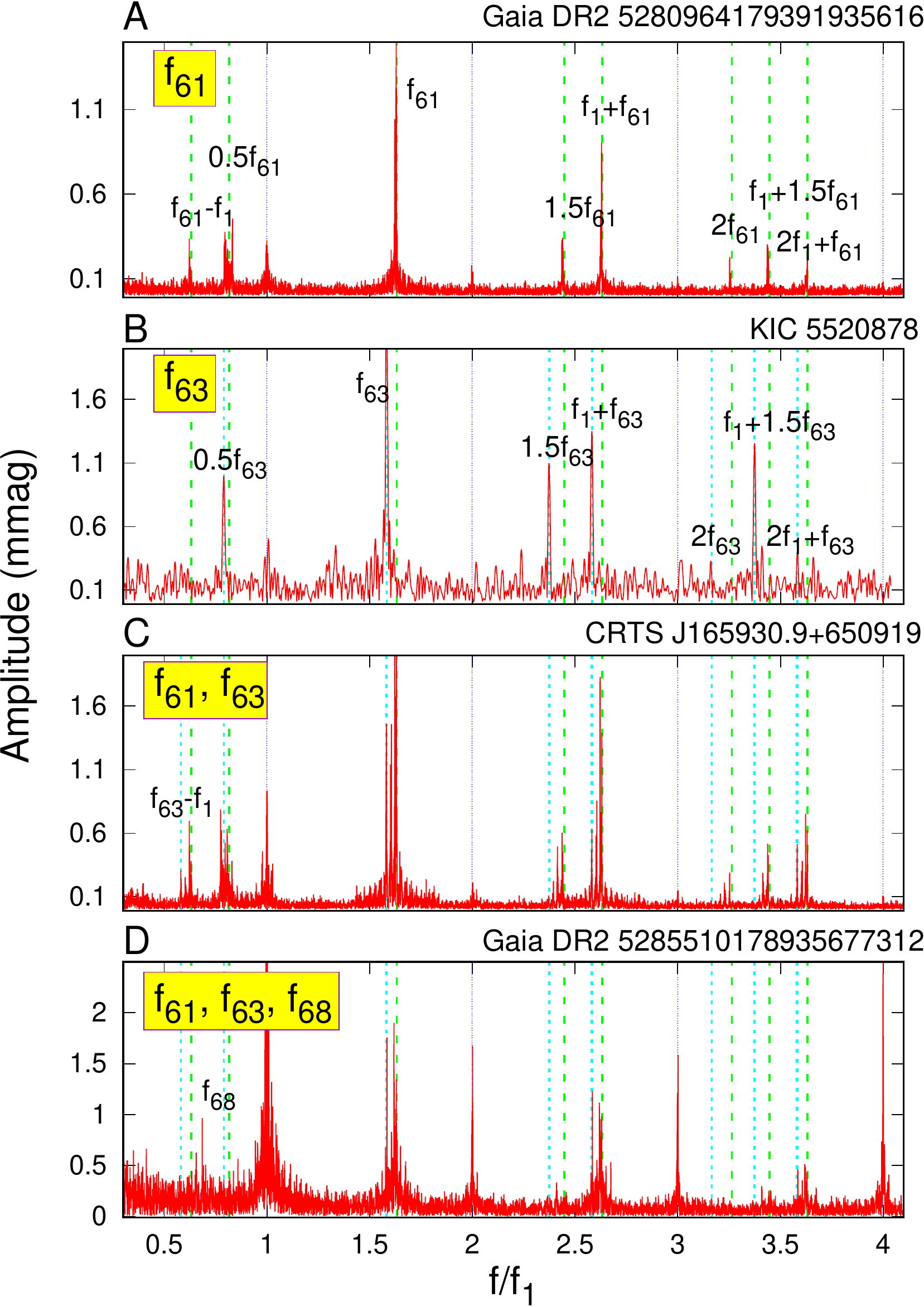}
\caption{Sample spectra of four stars where $f_{61}$ or $f_{63}$
is the dominant additional frequency. The spectra are pre-whitened with the overtone pulsation 
frequency $f_1$ and its harmonics.
The vertical blue lines show the locations of these pre-whitened components. 
Vertical green dashed and cyan dotted lines denote the positions of the 
identified and labeled frequencies in each panel.
For comparison, these vertical lines are redrawn from top to bottom 
on successive panels.
(See the text for the details.)}\label{fig:f61}
\end{figure}

The distribution in Fig~\ref{fig:zoom_hist} shows a local maximum
around 0.62. This must be caused by the frequencies with a 
ratio of $\sim0.622$ isolated by \citet{Netzel2019} 
in the OGLE material. As \citet{Molnar2022} has pointed out, 
in the mixed metallicity 
sample of the Galactic field, the individual ridges are less 
separated than in the case of the much more homogeneous Galactic 
bulge sample of OGLE. 
Nevertheless, all three ridges found earlier are clearly 
identified here, and their centre values coincide remarkably well 
with the values of stars from the Galactic bulge.
The frequencies around the ratio 0.62 are explained as linear
combination of the two above mentioned non-radial mode frequencies
viz. $f_{62}=0.5f_{61}+0.5f_{63}$.

In Fig.~\ref{fig:f61} we show four spectra as a sample in which 
the $f_{61}$ or $f_{63}$ frequencies are dominant. 
In the spectrum of panel A, the frequency of the mode $0.5f_{61}$, and its 
harmonics $f_{61}$, $1.5f_{61}$ and $2f_{61}$ are clearly visible. Some linear combinations of 
these frequencies with $f_1$ are also detectable.
The position of these detected frequencies are marked with vertical green dashed lines.
The spectrum in panel B is dominated by the $f_{63}$ frequency, 
otherwise the structure is very similar to the spectrum in panel A.
Here the identified peaks are marked with a cyan dotted line.
Panel C shows a spectrum in which both the $f_{61}$ and $f_{63}$ 
frequencies are present. The structure of the spectrum can be 
understood as a combination of the spectra of panels A and B. 
Between the frequencies $f_{61}$ and $f_{63}$, peaks belonging 
to the linear combination of the two modes, i.e. $f_{62}$, $f_{61}+0.5f_{63}$ also appear.
The star shown in the panel D features, in addition to 
the $f_{61}$ and $f_{63}$ frequencies, the $f_{68}$ 
frequency -- discussed in the next subsection -- as well.

\subsubsection{Stars with $f_{68}$ frequencies}

\begin{figure}
\includegraphics[width=0.45\textwidth, angle=0]{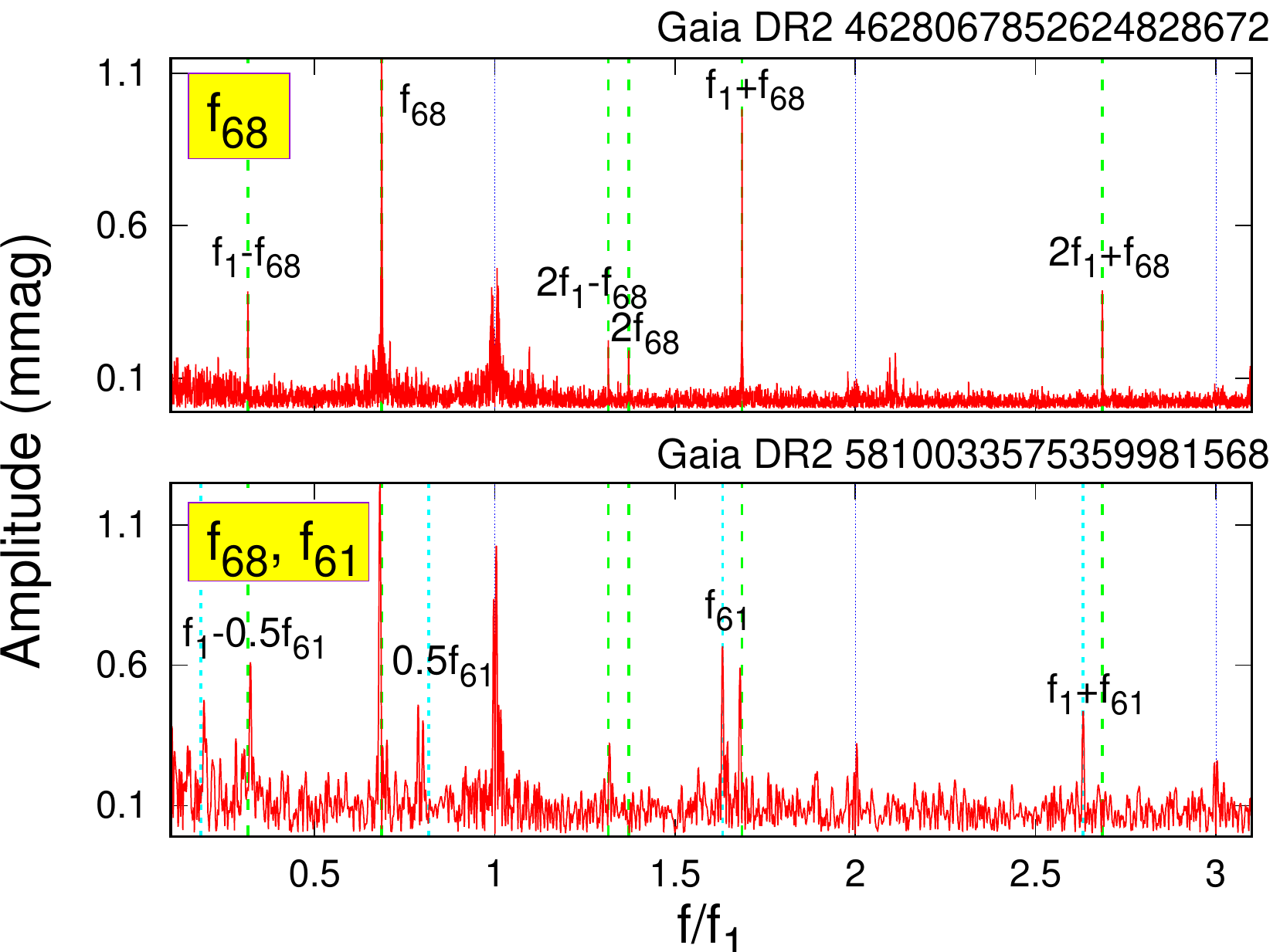}
\caption{Spectra of two stars where the $f_{68}$ signal
is the dominant additional frequency. 
The structure of the figure is the same as Fig~\ref{fig:f61}.}
\label{fig:f68}
\end{figure}

Previous works on overtone RR Lyrae stars (\citealt{Netzel2015,Moskalik2015,Molnar2015})
and Cepheids \citep{Suveges2018, Smolec2023} also identified an independent additional frequency with 
a relatively large amplitude that is present in many stars. 
These frequencies have a ratio $P_x/P_1=1.459$ or $P_1/P_x=0.685$
(hereafter signed as $f_{68}$).
Based on analogies, \citet{Moskalik2015} has suggested non-radial g-modes as a possible origin of these low frequencies, but no theoretical modeling has been done on this hypothesis. 
\citet{Dziembowski2016} has investigated another possible explanation 
for $f_{68}$, namely that the stars 
showing these are not RR Lyrae variables, but 
low-mass binary `RR Lyrae impostors', or binary evolutionary pulsators,
such as OGLE-BLG-RRLYR-02792 \citep{Pietrzynski2012, Karczmarek2017}.
However, he rejected this possibility since
such an explanation is contradicted by the 
existence of stars with $f_{68}$ together 
with $f_{61}$ frequencies (e.g. KIC\,9453114 in the \textit{Kepler} field \citealt{Moskalik2015}, 
ASAS\,J213826-3945.0 and NSVS\,14632323 in \textit{TESS} data \citealt{Molnar2022}).
Most recently, \citet{Molnar2022} mentioned, but only 
as an unlikely possibility, a heavily damped $\ell=1$ mode that causes these frequencies.

The (green) horizontal ridge of $f_{68}$ frequencies is 
clearly visible in Fig.~\ref{fig:Petersen}. Spectra of 133 stars (21 per cent of the sample)
contain frequencies with this period ratio.
Along with these, we also found a number of linear combination
frequencies that contain the $f_{68}$ frequency (see Fig~\ref{fig:f68} and Table~\ref{tab:fr_track}). 
The presence of linear combination frequencies with the radial mode
frequencies (even in flux data) reinforces the picture that $f_{68}$ 
signals also belong to RRc stars (and not e.g. to a companion star).

Generally, the highest amplitude linear combination frequency
of these is $f_1+f_{68}\equiv f_{59}$.
These frequencies cause the sharp peak at  $P_x/P_1\sim 0.593$ on 
the left side in Fig.~\ref{fig:zoom_hist}. 
Previous publications have not reported similar 
frequencies, however, \citet{Dziembowski2016} shows that the 
frequencies corresponding to the $\ell=10$ non-radial mode 
should be located around here.
Is it possible that this is the true mode frequency 
(precisely half of it) and the mysterious $f_{68}$ 
frequency is just a simple linear 
combination (i.e.: $f_{68}=f_{59}-f_1$)?

A search of the Fourier spectra shows that $f_{59}$ frequencies are 
only found in those 67 stars where $f_{68}$ also appears. 
This strengthens the connection between $f_{59}$ and $f_{68}$.
In all such cases, $f_{59}$ has a lower amplitude than $f_{68}$. 
This suggests that $f_{68}$ is the primary frequency and $f_{59}$ 
is the linear combination. Although in some special cases, the linear combination 
frequencies of non-radial modes may have higher 
amplitudes than their component frequencies \citep{Balona2013, Kurtz2015}, this is not true 
for linear combinations with radial modes.
This fact is against interpreting the $f_{68}$ frequency as 
a linear combination. 
A further argument against the primarity of $f_{59}$ is that 
none of the stars containing $f_{59}$ has a frequency 
corresponding to $0.5f_{59}$ (the hypothetical frequency of $\ell=10$ mode).

For at least five stars, the harmonic frequency $2f_{68}$ was detected 
(see top panel of Fig~\ref{fig:f68} for an example).
The appearance of higher-order, even fifth-order, harmonics of frequencies in certain stars belonging to the $\ell=8$ and $\ell=9$ non-radial modes indicates that non-linear effects 
also play role in the excitation of these modes (see e.g. \citealt{Cox1980}). 
Something similar is suspected from the appearance of $2f_{68}$ harmonic frequency.

We found 66 such stars with at least one additional frequency (usually $f_{61}$) besides $f_{68}$.
What is interesting, however, that linear combinations of $f_{68}$ frequencies 
with $f_{61}$ or $f_{63}$ frequencies were not found for any stars, even though $f_{62}$ 
(a linear combination of $f_{61}$ and $f_{63}$) is so common that it causes a clearly visible ridge in the diagram shown in Fig.~\ref{fig:Petersen}.
In stars where both $f_{68}$ and $f_{61}$ are present, 
$f_{61}$ is almost always dominant, i.e., the spectrum is similar 
to that in panel D of Fig.~\ref{fig:f61}. There are only eight stars 
where $f_{68}$ is dominant for at least two sectors (see bottom panel of
Fig.~\ref{fig:f68} for an example). The latter 
is important because the amplitude of the $f_{61}$ frequency is 
strongly time dependent (see \citealt{Szabo2014, Moskalik2015} 
and later in Sec.~\ref{sec:t_add}). 
There are several stars where $f_{68}$ dominates 
in one sector but $f_{61}$ dominates in the following sector or vice versa.

\subsubsection{Frequencies of the $\ell=10$ modes?}\label{sec:f60}

\begin{figure}
\includegraphics[width=0.45\textwidth, angle=0]{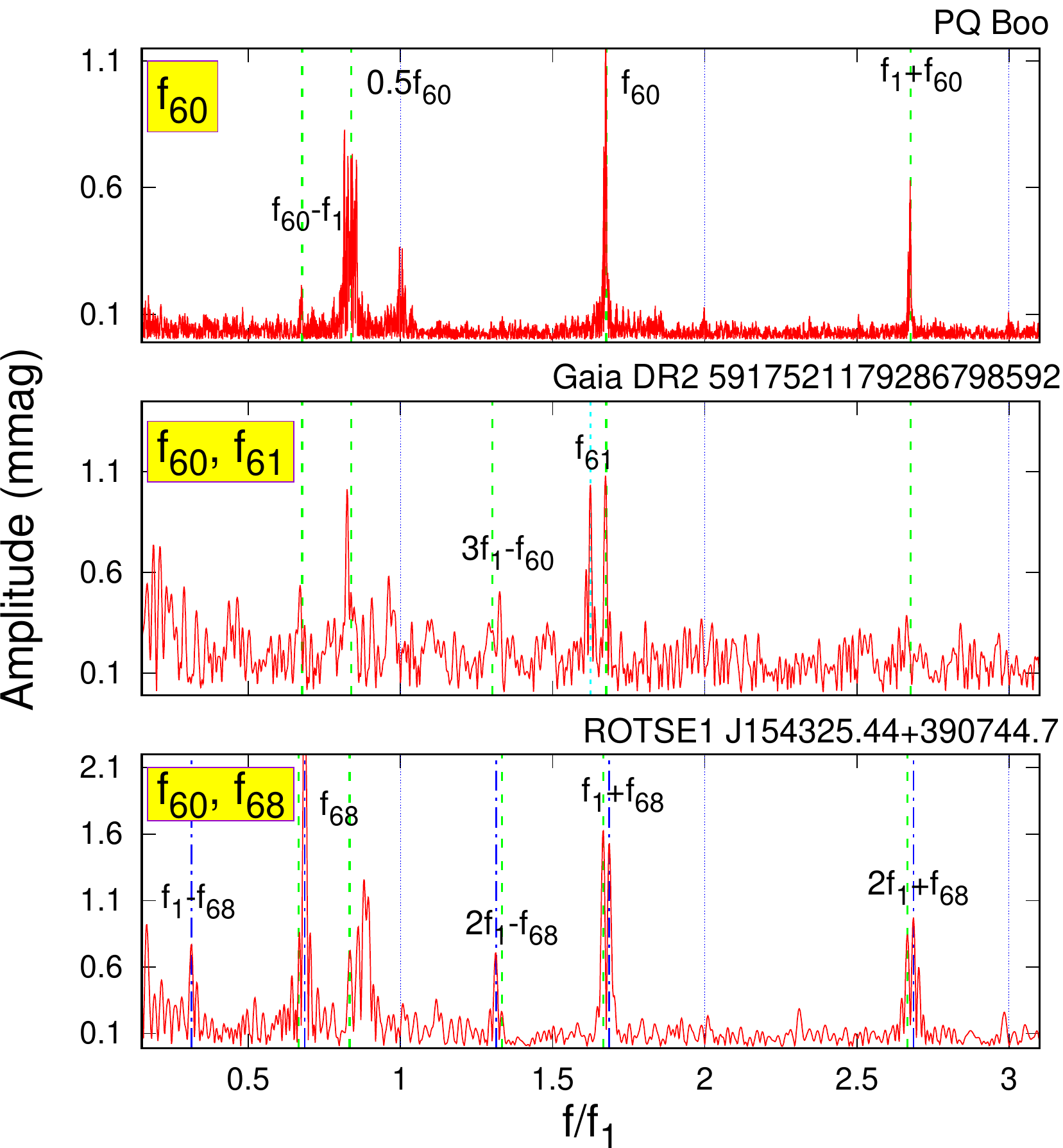}
\caption{Spectra of stars showing $f_{60}$
frequency and its linear combinations (top);
the $f_{60}$ and $f_{61}$ frequencies (middle); or $f_{60}$ and $f_{68}$ frequencies (bottom), simultaneously.
The structure of the figure is the same as Fig~\ref{fig:f61}.}\label{fig:f60}
\end{figure}
The histogram in Fig.~\ref{fig:Petersen} suggests that 
in addition to the 
four frequent period ratios discussed above (associated
with frequencies $f_{61}$, $f_{63}$, $f_{62}$ and $f_{59}$), 
one more period ratio is relatively frequent in this interval, 
around 0.602. 
A frequency with a period ratio between 0.595 and 0.605 was 
found in 97 stars, 21 
of which are the dominant additional frequency.
(Hereafter, the frequencies corresponding to this ratio are 
denoted by $f_{60}$.)

If we look at the Fourier spectra of stars with these frequencies
(see top panel in Fig.~\ref{fig:f60} for an example),
we find that they are similar to the spectra of stars with 
$f_{61}$ frequencies. 
Typically, the $0.5f_{60}$ frequency is also detectable, as well 
as various linear combinations of the $f_{60}$ frequency and $f_1$.
One could think that these are also $f_{61}$ stars, but for some reason their period ratio is unusually small.

However, this assumption certainly cannot be applied to stars for which the frequencies $f_{60}$ and $f_{61}$ appear simultaneously.
The middle panel of Fig.~\ref{fig:f60} shows the spectrum of such a star. The frequencies 
associated with $f_{60}$ and $f_{61}$ are clearly separated.
For  Gaia\,DR2\,5917521179286798592, shown in the figure, 
the difference between the two frequencies is 
$\Delta f=f_{60}-f_{61}=0.185$~d$^{-1}$. 
This is more than five times higher than the Rayleigh frequency resolution (0.036~d$^{-1}$). 
Frequency splitting might be caused by either 
amplitude or frequency modulation \citep{Benko2011}. 
This is, however, not the case here. The main pulsation is rather stable, 
as indicated by the small residuals around the harmonics. 
Using the amplitude and phase variation calculating
tool of the {\sc Period04} package, we also tested the stability of the $f_{61}$ 
amplitude and phase. We divided the light curve into 2-day slices and determined 
the phase and amplitude of the same trial frequency for each slice. 
We found that the phase is constant to a good approximation
($0.0179\pm0.0045$~rad), while the amplitude shows a 
wave with a timescale of $\sim15$ days and amplitude of $0.75$~mmag. 
This variation could explain at most a side peak at a distance of 0.067~d$^{-1}$. 
In other words, neither the time variation of the 
main nor the additional frequency can explain the 
$f_{60}$ frequency appearing next to $f_{61}$.

Since the $f_{59}=f_{68}+f_1$ frequency is close to the 
$f_{60}$ one, we also need to show that the $f_{60}$ stars are different from $f_{68}$ stars.
There are only a few (3--4) candidates
where we may assume that the $f_{59}$ and $f_{60}$ frequencies appear simultaneously.
The best candidate is shown in the bottom 
panel of Fig.~\ref{fig:f60}. In this case, $\Delta f=f_{59}-f_{60}=0.06$~d$^{-1}$, which 
coincides with the instrumental variations that might be caused by the orbital period.
This is not possible, however, because in that case the whole light curve would be affected.
Additionally, the instrumental origin does not explain 
the peaks between $f_1$ and $f_{68}$, which also appear 
in all the other similar stars and can naturally 
be explained by a time-varying and therefore multi-peaked 
frequency at $0.5f_{60}$. This type of 
`composite' spectrum also appears for simultaneous
$f_{61}$ and $f_{68}$ stars (bottom panel in Fig~\ref{fig:f68}).

\begin{figure}
\begin{center}
\includegraphics[width=0.24\textwidth, angle=270]{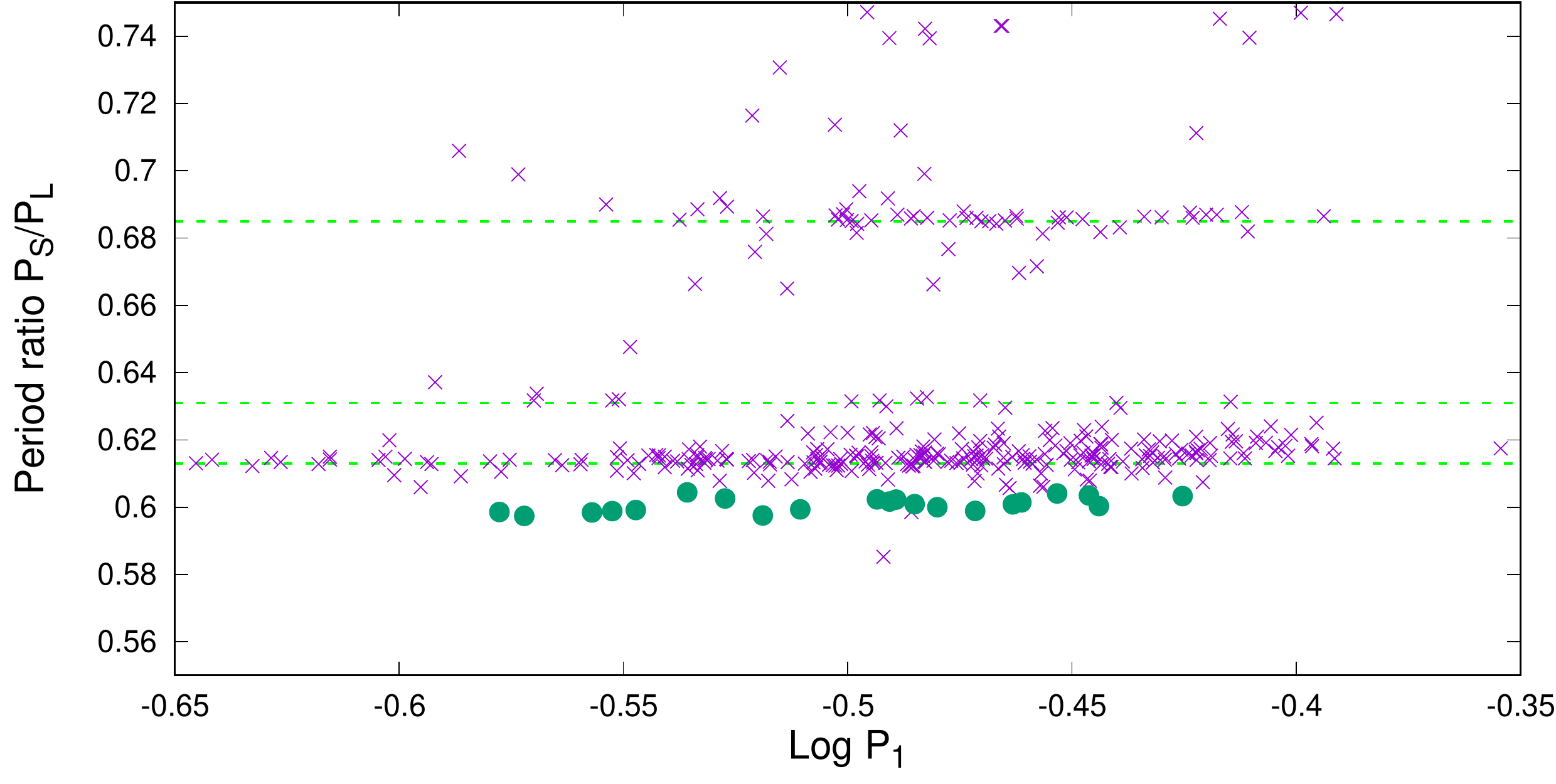}
\includegraphics[width=0.47\textwidth]{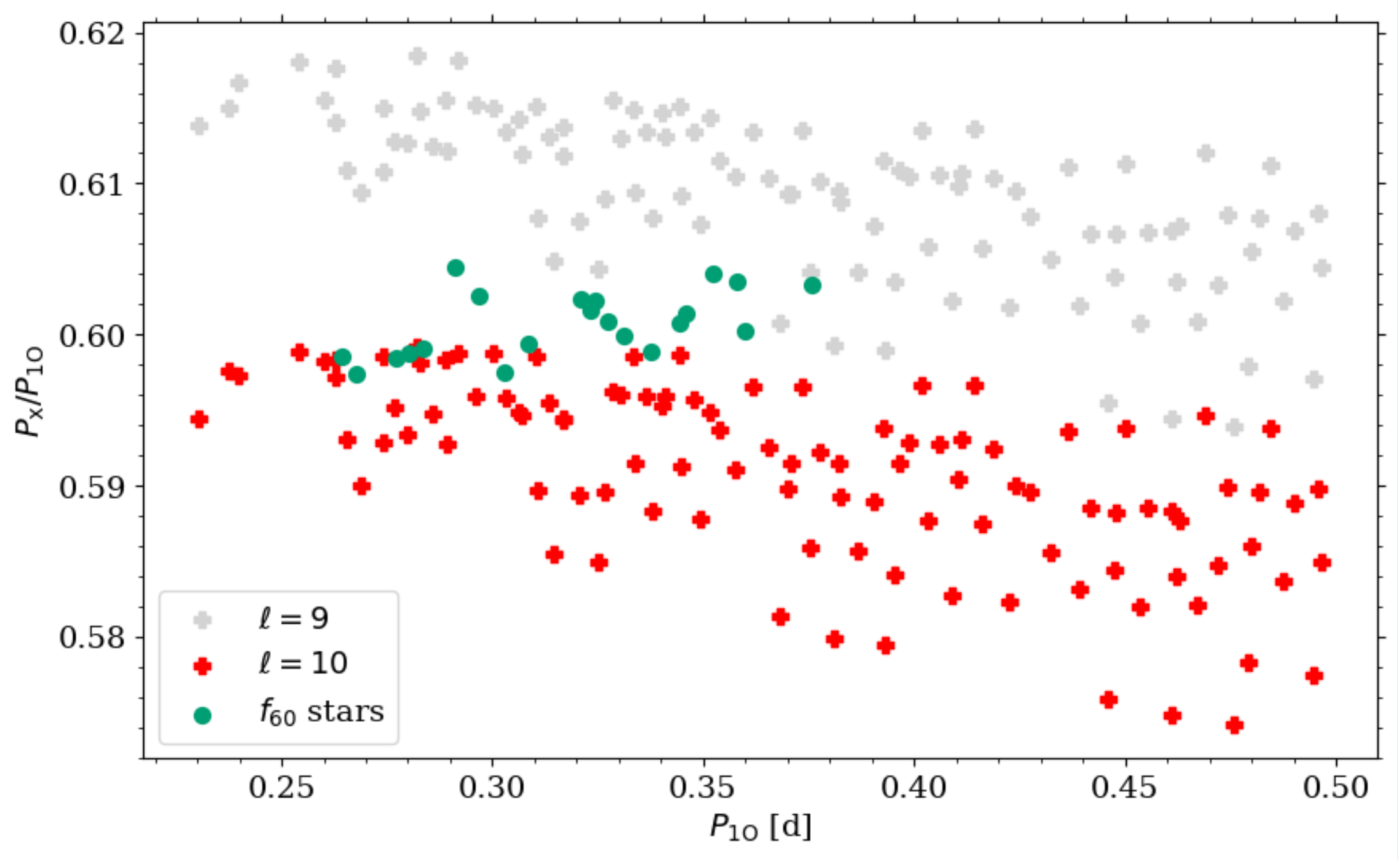}
\end{center}
\caption{(top) Part of the Petersen-type diagram.  
Each symbol represents a star with its dominant additional frequency. Dashed horizontal lines indicate the middle position
of the known ridges at 0.613, 0.631 and 0.685, respectively.
The stars of the new $f_{60}$ ridge are shown with filled green circles.
(bottom) Theoretical models for RR Lyrae stars with non-radial modes of degrees 9 and 10 (see text for details). The stars of the new $f_{60}$ ridge are shown with filled green circles.} 
\label{fig:f60_ridge}
\end{figure}

In the upper panel of Fig.~\ref{fig:f60_ridge} we plotted the 
dominant additional frequencies of each star. 
As can be seen, stars containing 
$f_{60}$ (green filled circles) are quite distinct  
from the previously 
identified $f_{61}$ group (purple `x' symbols).
There are two possible explanations for this $f_{60}$ group. 
On the one hand, it is possible that the stars 
that make them up define a metal-poor subgroup of RRc stars. On the other hand, the frequencies associated with the 
$\ell=10$ non-radial modes are expected to be somewhere around 
$f_{60}$ (see \citealt{Dziembowski2016}).

In the bottom panel of Fig.~\ref{fig:f60_ridge} we compared the position of the $f_{60}$ group on the Petersen-like diagram with theoretical models of RR Lyrae stars with non-radial modes of degrees $\ell=9$ (blue symbols) and $\ell=10$ (red symbols). Theoretical models were calculated with the Warsaw envelope code \citep{dziembowski1977} for different values of mass, metallicity, luminosity and effective temperature. The ranges of these parameters are the same as in Netzel et al.\ (in prep.). We plotted only those models in which both the radial first overtone and the non-radial mode of a given degree are linearly unstable. To plot the period ratios we used the first harmonic of the non-radial mode, similarly to Fig.~4 by \cite{Dziembowski2016}. Positions of short-period $f_{60}$ stars are consistent with theoretical models of non-radial modes of degree 10. The $f_{60}$ stars with first-overtone periods longer than around 0.32\,d are located between the sequences formed by the theoretical models for non-radial modes of degrees 9 and 10. 
For these stars, 
linear combinations of $\ell=9$ and $\ell=10$ mode frequencies
might be considered, in a similar way as we explained $f_{62}$ frequencies 
by linear combinations of $\ell=8$ and $\ell=9$ mode frequencies.

The models therefore indicate that the $f_{60}$ ridge at shorter periods cannot 
be part of the $f_{61}$ ridge, but the explanation of the 
$\ell=10$ ridge in the non-radial mode is plausible. 
We note that the observed ratios for both $f_{61}$ and $f_{63}$ slightly increase with increasing period, while the models suggest an opposite change.

\subsubsection{Further additional frequencies}

\begin{figure}
\includegraphics[width=0.35\textwidth, angle=270]{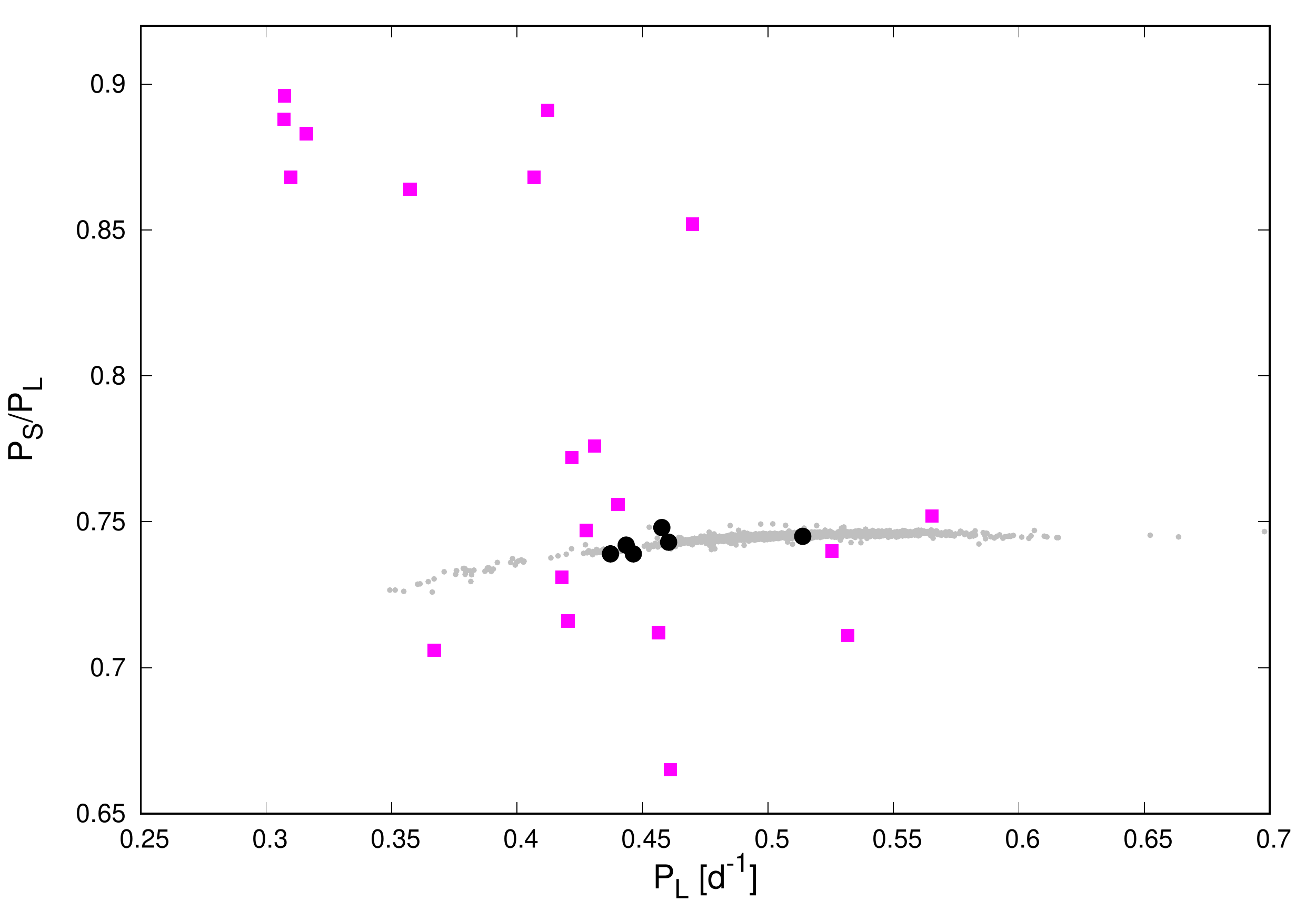}
\caption{Stars with a dominant additional period 
ratio $P_S/P_L=P_1/P_{\mathrm x}$ between 0.65 and 0.9. 
%Filled black circles indicate the possible RRd stars.
Grey dots show stars of the OGLE RRd sample \citep{Soszynski2014}.}\label{fig:rrd}
\end{figure}

We also investigated what other frequencies occur in addition to those discussed above.
The classical Petersen diagram in Fig.~\ref{fig:rrd} shows the 26 stars 
whose dominant additional frequencies do not fit into the previous categories.

Two groups appear to be distinct: one between period ratios 0.665 and 0.776 and another one between 0.85 and 0.896.
The first case also includes stars that fall into the sequence of classical RRd stars (black circles),
i.e., for these six stars the dominant additional frequency found could be the frequency of the fundamental mode, $f_0$. In Table~\ref{tab:Four_res} these stars are denoted by `$f_0$?'.
Two aspects, however, seem to contradict the classification of these stars as RRd variables.
(i) For these stars, the amplitude ratio of the 
two frequencies $A_0/A_1$ is between 0.032 and 0.019. 
The smallest such value for classical RRd stars is an order of magnitude larger. 
(ii) The $f_0$ frequencies typically appear among a broad forest of peaks. This is the case for all frequencies shown in Fig.~\ref{fig:rrd}. 
The amplitudes and frequencies of the additional frequencies vary over time. This is particularly 
true for the frequencies of the modes themselves (i.e. $0.5f_{60}$, $0.5f_{61}$, $0.5f_{63}$). The consequence of this is the appearance of side peaks (see Figs.~\ref{fig:f61} and \ref{fig:f60}). The amplitude of the side peaks, which may be larger than the main peak, is determined by the actual time dependence \citep{Benko2011}.

To summarize: although the frequency ratios calculated from the doubles of the 
frequencies discussed here do not match the canonical values of the $f_{6x}$ 
ratios, they are always close to them. Additionally, a bunch of peaks are near them, thus
they are likely to be one of the extreme (right or left) side peaks 
due to the time dependence of the mode frequencies $0.5f_{6x}$.
That is, the two groups in Fig.~\ref{fig:rrd} are in fact apparent.

It should be noted that these 
stars are not the same as the so-called anomalous RRd stars \citep{Soszynski2016aRRd}, since for all anomalous
RRd stars the $f_0$ frequency is dominant compared to $f_1$.  
Furthermore, there is only a marginal overlap between these and the new group 
described by \citet{Prudil2017}, as they have $P_{\mathrm L} < 0.4$~d.

\subsubsection{Additional frequencies and the period dependence of Fourier parameters}\label{sec:per_distr}

\begin{figure}
\includegraphics[width=0.3\textwidth, angle=270]{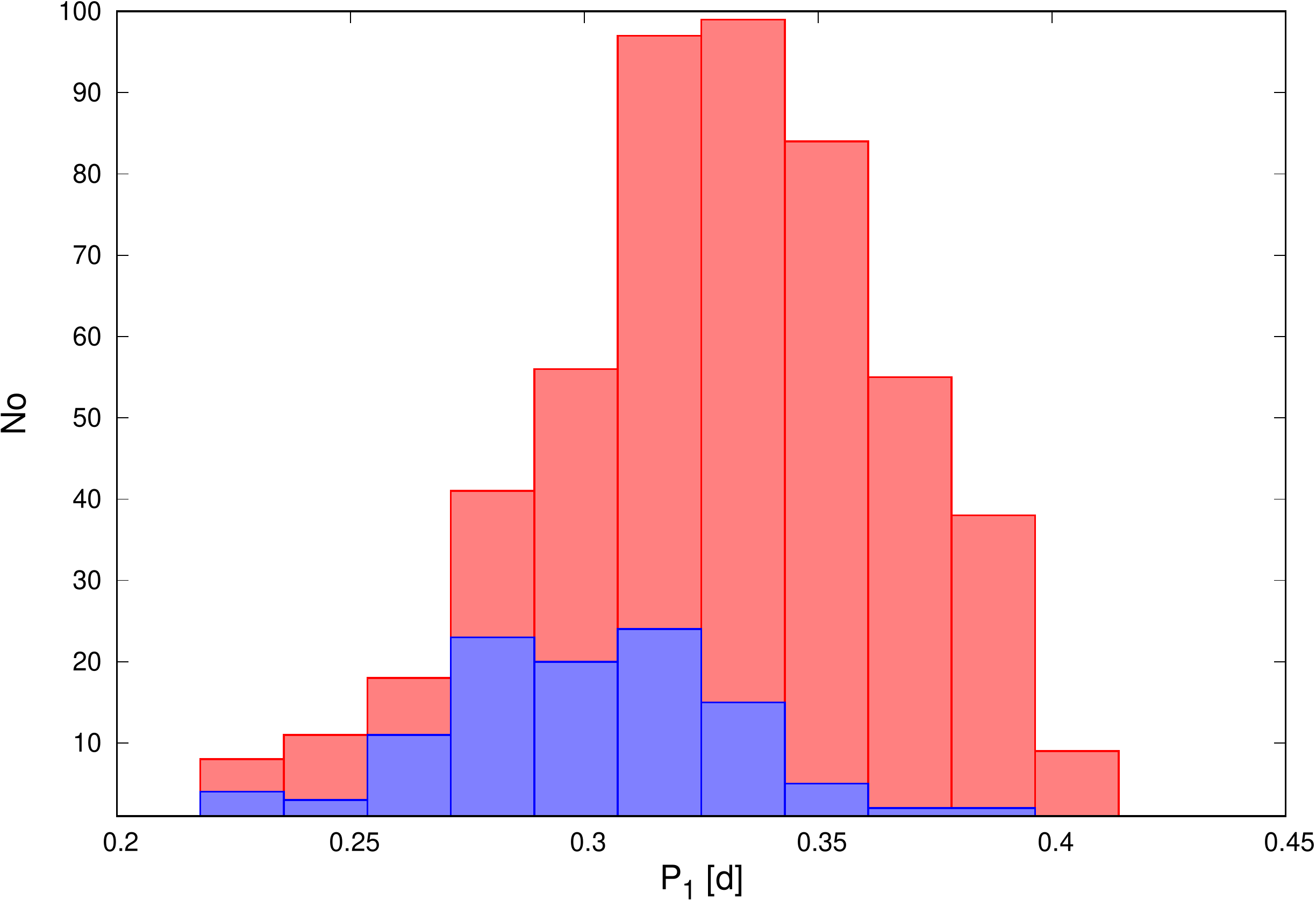}
\caption{
The number of stars that show show additional frequencies (red boxes) and those that do not (blue boxes),
as a function of the main period.
}\label{fig:hist+-}
\end{figure}
Fig.~\ref{fig:Fourier_param} shows not only that the Fourier parameters of our sample coincide 
with those of the RRc stars, but also that the parameters of the stars with additional frequencies 
(red symbols) and those without (blue symbols) are somewhat different.
This indicates that there might be a physical difference between the two groups, and that the non-detection of 
additional frequencies in certain stars does not happen only for observational reasons.

The obvious feature is that while the stars with extra frequencies occur over the whole parameter range, almost all stars without extra frequencies fall into the shorter period ($P<0.35$~d) range. 
The phenomenon is illustrated by the distribution of stars in Fig.~\ref{fig:hist+-}. 
In each period bin, stars with detectable additional frequencies are the majority, but stars that 
do not show such frequency are concentrated in the shorter periods and almost absent in the longer period bins. Or, in other words, the ratio of those stars that do not show an additional frequency to those ones that do decreases as the period increases.
\begin{figure*}
\includegraphics[width=0.33\textwidth, angle=270]{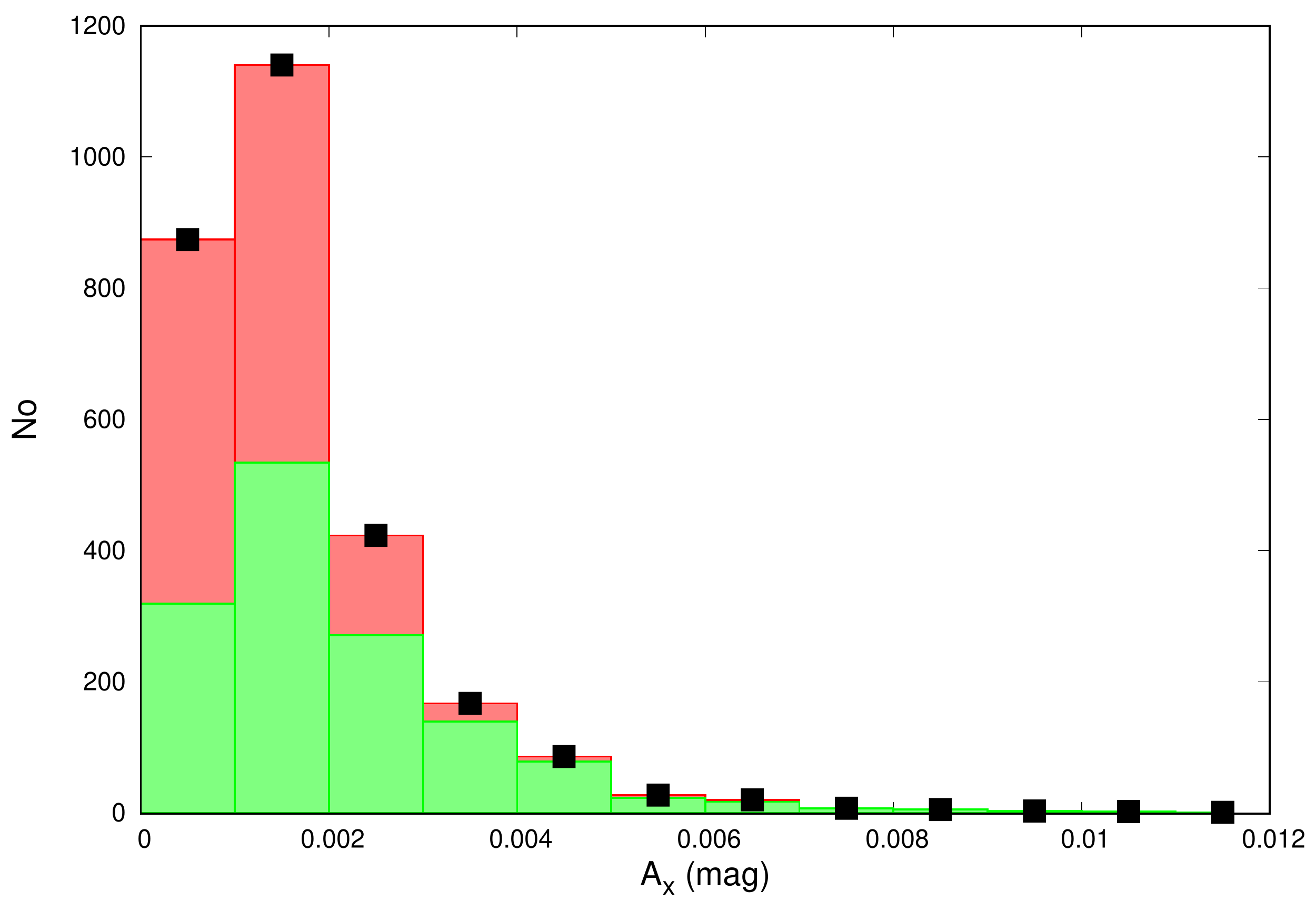}
\includegraphics[width=0.33\textwidth, angle=270]{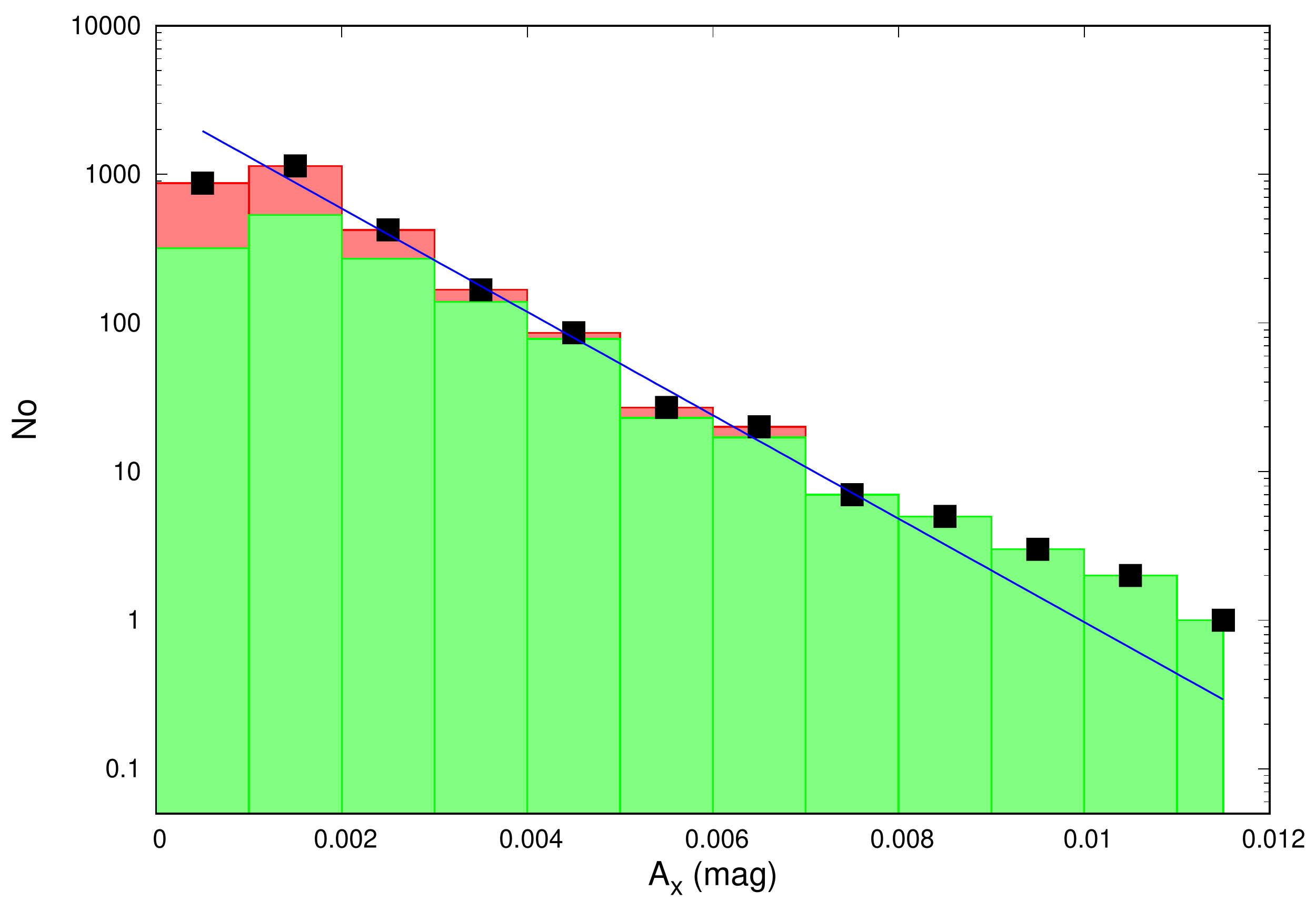}
\caption{Amplitude distribution of 0.61 additional modes. 
The number of found additional frequencies as a function of their amplitudes in normal (left) and in logarithmic (right) scales.
Green columns show the frequencies in the 0.61 sequence, while 
red ones indicate all frequencies between $P_{\rm S}/P_{\rm L}=0.55$ and 0.65.
\label{fig:amp_distr}}
\end{figure*}

Recently, \citet{Netzel2022} have derived theoretical models to determine the parameters under which the $\ell=8$ and $\ell=9$ non-radial modes can be excited in the overtone pulsating RR Lyrae stars. 
Excitation of these non-radial modes is possible over a wide range of metallicities and over 
most of the instability strip. However, as the blue edge of the instability strip is approached, the excitation of the $\ell=8$ and then the $\ell=9$ non-radial modes are no longer present either
(see fig.~2 in \citealt{Netzel2022}).
This means stars that do not excite non-radial modes are more likely to be among the hottest RRc stars. 
This phenomenon also manifests in the fact that if the possible 
$\ell=8$ or $\ell=9$ regions are plotted on the Petersen diagram (see fig.~3 in \citealt{Netzel2022}), they are wedge-shaped: longer periods allow a wider period 
ratio for the excitation of these non-radial modes than shorter periods. 
Comparing this with what we found from the observations, we can say that the 
theoretical prediction and the measurements, at least at a qualitative level, gave similar results.

That stars without an additional mode are more frequent among shorter-period RRc was 
shown not only by 
the smaller \textit{TESS} sample of \citet{Molnar2022}, but also by the stars of \citet{Jurcsik2015}
in the globular cluster M3, but it became evident with the present larger \textit{TESS} sample. 
The phenomenon supports that the theoretical explanation for the additional frequencies is correct.

\subsubsection{Amplitude distribution}

For comparison, the amplitude distributions of the additional frequencies 
were calculated in a similar way as done by \citet{Netzel2019}:
we considered the frequencies either between 0.6 and 0.62 ($f_{61}$ stars), 
and also the frequencies between 0.55 and 0.65.
We plotted in Fig.~\ref{fig:amp_distr} the number of additional frequencies found as a function of their 
amplitude for these two cases. In both cases the distributions are very similar to each other and 
to the similar plots prepared to the OGLE sample (see fig. 4 and fig. 5. of \citealt{Netzel2019}).
The red boxes disappear at high amplitudes in Fig~\ref{fig:amp_distr}. 
This means that $f_{63}$ have typically lower amplitudes than $f_{61}$ frequencies.
For values larger than 0.01~mag amplitude, the distribution 
appears similar to an exponential one, while for smaller amplitudes, 
there is a deficiency compared to the exponential distribution.
The relatively smaller number of frequencies with small amplitudes would be 
understandable at first glance, since the smaller the amplitude, 
the less likely we are to detect them.

However, the logarithmic scale plot shows another difference from 
the exponential distribution: we detect more frequencies with large amplitudes 
than we would expect from such an exponential distribution. 
To illustrate this phenomenon, we constructed a data series from our
 histograms (black rectangles in Fig.~\ref{fig:amp_distr}). The central part 
of the logarithmic version of the data set (between $A_x=0.0015$ and 0.008~mag) was linearly 
fitted (blue line). The fit of the line is excellent (rms=0.18), towards 
higher amplitudes the excess in the number of detected frequencies is evident.

Using `FindDistribution' tool of the {\sc Mathematica} program 
package \citep{Mathematica12}, we searched for 
the distribution that best fits our data set. The solution is, however, the `empirical' distribution. 
The Zipf distribution (see e.g. \citealt{Johnson2005}) suggested as second possibility gives
worse fit than the above exponential one.
If we search for a distribution by adding to the first element of the histogram the 
109 stars in which no additional frequencies were found, i.e. assuming that the additional frequencies are of such small amplitude that we could not detect them, 
the result is the same.
All we can say is that the amplitude distribution of the additional frequencies 
is exponential over a wide range, but there is a deficit for the lowest amplitudes 
and a surplus for the highest amplitudes.

\subsubsection{Distribution in space}

\begin{figure}
\includegraphics[width=0.33\textwidth, angle=270]{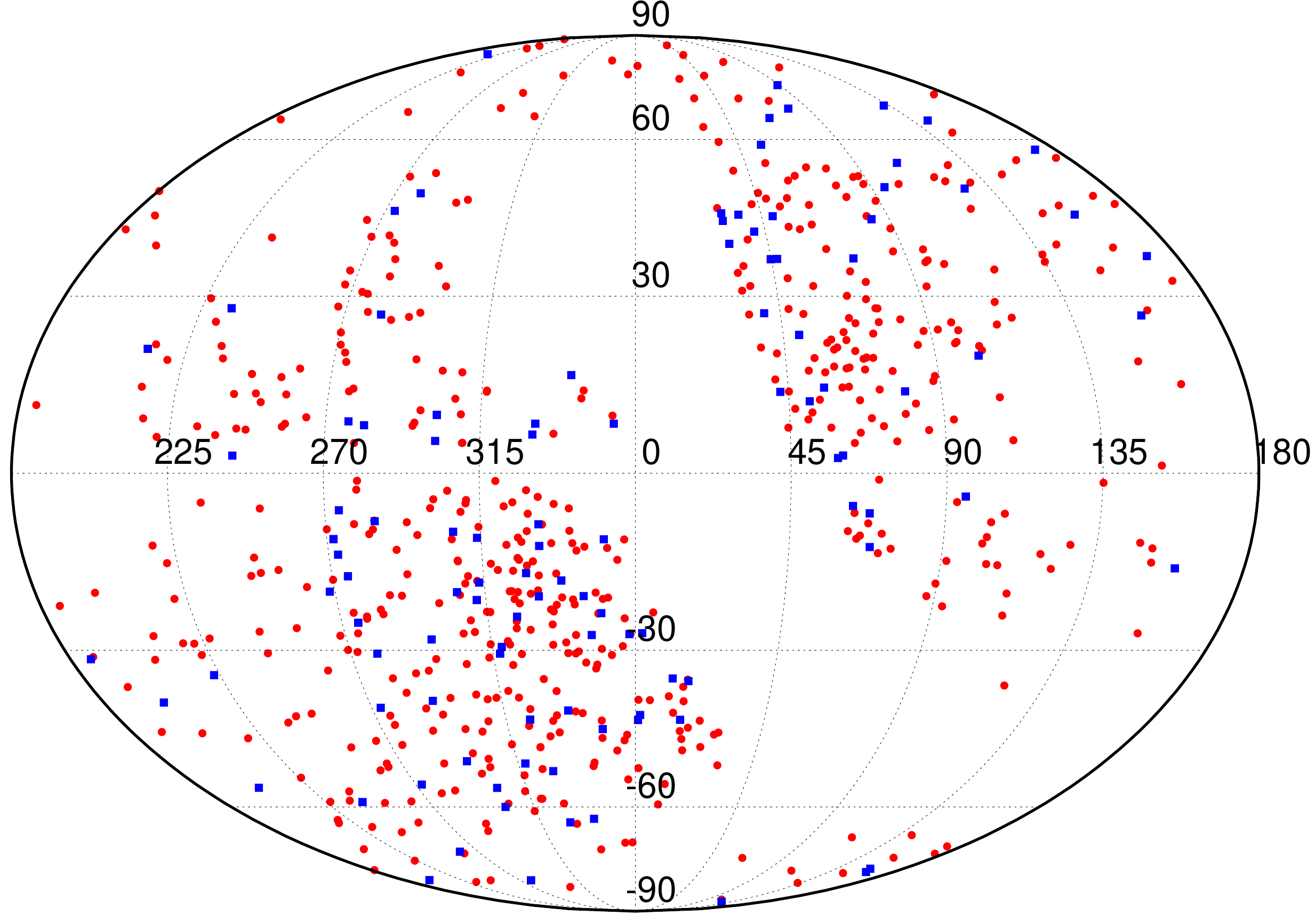}
\caption{Galactic 
distribution of the bright \textit{TESS} 
RRc sample. Red dots: stars with additional modes,
blue rectangles: stars without additional modes.
\label{fig:Galactic_distr}}
\end{figure}
% A table_1.gal+ es table_1.gal-
% alapjan a spherical_Mollweide.pt gyartja le. 
% A vegleges adatokra ujracsinalva!! -- OK
In subsection \ref{sec:per_distr} we saw that stars that do not exhibit the additional mode are 
different from those that do: they have typically shorter periods.
What is the distribution of these stars in the Galaxy?

In Fig.~\ref{fig:Galactic_distr}, we have plotted our 
stars in galactic coordinates. The red dots represent 
stars that show the additional mode(s), while the 
blue rectangles show stars that do not.
We examine whether there is a difference in the 
spatial distribution of stars with and without additional modes. 
To this purpose, a two-dimensional two-sample 
Kolmogorov--Smirnov (KS) test \citep{Press1988}
was applied to the coordinates of the two sets of stars. 
The two-sample KS test does not make any assumptions about 
the actual shape of the distributions, but only measures 
whether or not the two data distributions can be from 
the same distribution. As \citet{Press1988} has shown 
if the test result probability $p_r$ is larger than 0.20
then the assumption that the two data distributions are not significantly different is certain. 
In our case $p_r=0.487$, so we can conclude that the spatial distributions 
of RRc stars with and without additional 
modes do not differ.

\subsubsection{Distribution of additional frequency types}\label{sec:distr_add_types}

\begin{figure}
\begin{center}
\includegraphics[width=0.37\textwidth, angle=270]{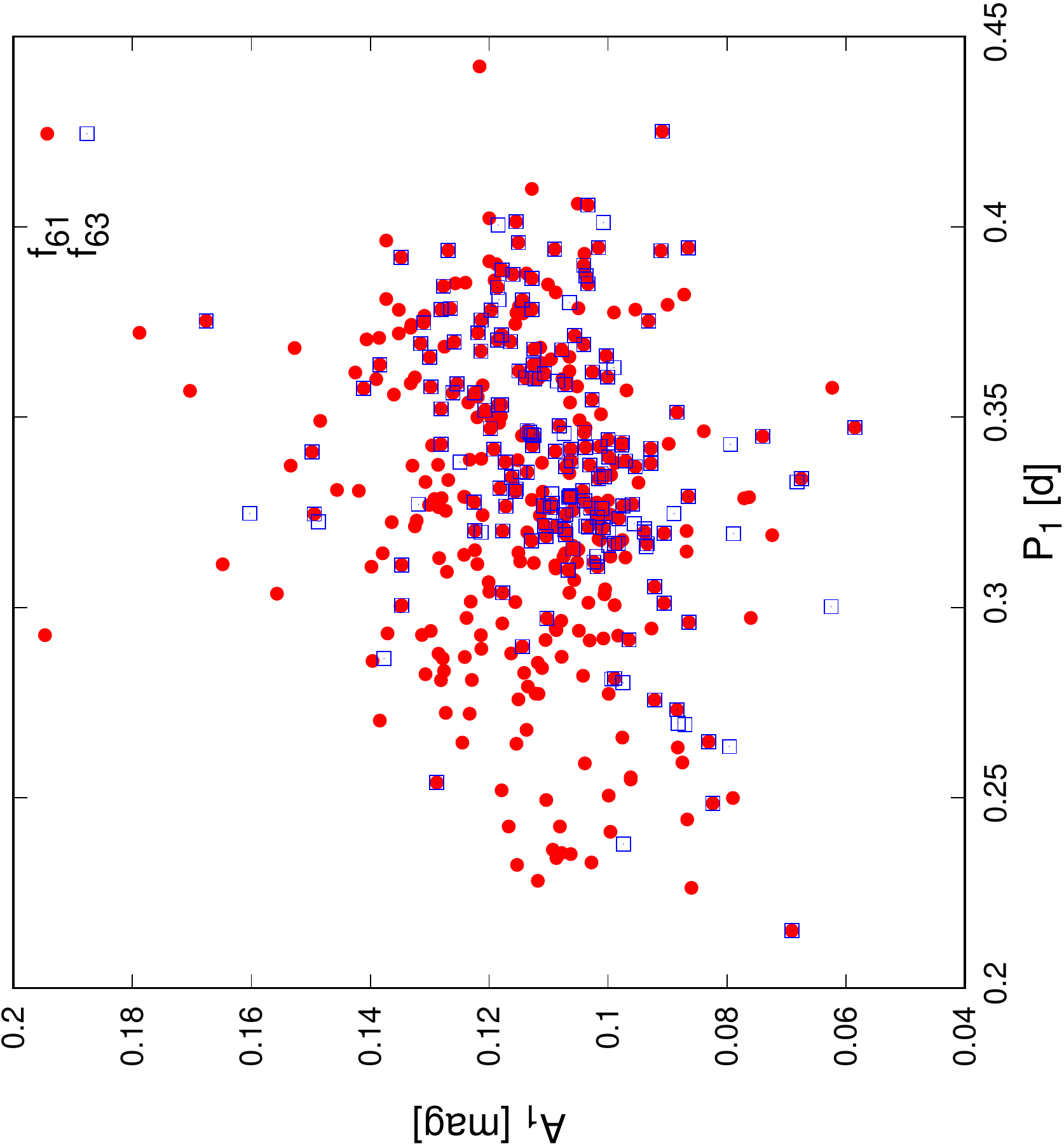}
\includegraphics[width=0.37\textwidth, angle=270]{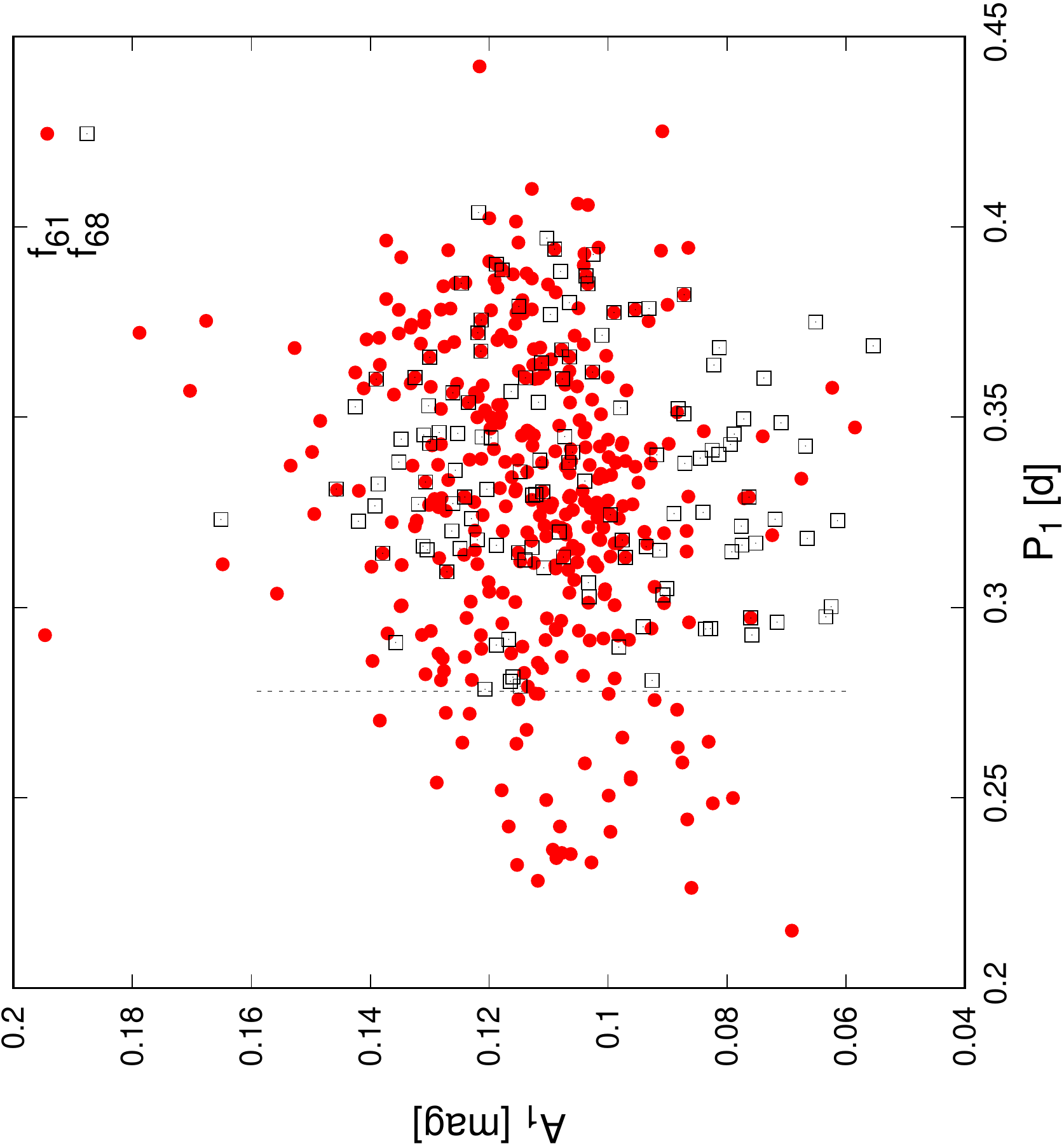}
\end{center}
\caption{Appearance of the $f_{61}$ and $f_{63}$ additional frequencies as a function of $P_1$ (top).
Filled red circles and empty blue rectangles show the positions of 
stars which have either $f_{61}$ or $f_{63}$ frequencies. 
Many stars show both. Stars with  $f_{61}$ and  $f_{68}$ frequencies (bottom).
Red points are the same as above, while empty black rectangles represent stars
with $f_{68}$ frequencies.
}\label{fig:adist}
\end{figure}
In Sec.~\ref{sec:per_distr} we saw that stars that do and do not show an 
additional frequency, regardless of its type, 
are distributed differently on the period vs. Fourier parameter planes.
We also saw in Sec.~\ref{sec:f60} that the period distribution $f_{60}$ stars appears to be relatively even. 
In any case, the number of elements in our $f_{60}$ sample is too small to make any strong statements.
What about the distribution of other types? 

In the top panel of Fig.~\ref{fig:adist} we plotted the stars showing $f_{61}$ (filled red circles) 
and $f_{63}$ (empty blue rectangles) additional frequencies as a function of $P_1$. 
The vertical scales show the amplitude of the main period $A_1$.
An evident structure can be recognized in the figure: stars with shorter periods and higher
amplitudes typically show only $f_{61}$, while stars with lower amplitudes and
longer periods typically show both $f_{61}$ and $f_{63}$ frequencies.
If we look at the theoretical HRD of \citet{Netzel2022}, we see that there is indeed a range 
where only the $\ell=9$ non-radial mode can be excited, but the $\ell=8$ cannot. 
This range is, however, rather narrow. 
As shown by \citet{Bellinger2020} the amplitude of the light curve and the period
are the two most important parameters that determine the effective temperature of RR Lyrae stars.
However, this is only a correlation, not a simple function, since luminosity, mass and chemical composition also play a non-negligible role
(see e.g. \citealt{Marconi2015} and further references therein).
To make a quantitative comparison between the theory and observation would require a 
deeper analysis, which is beyond the scope of this paper.

In the bottom panel of  Fig.~\ref{fig:adist} we show with red filled circles 
the same $f_{61}$ stars as  
in top panel while the black empty squares denote the stars containing $f_{68}$ frequencies.  
A sharp threshold is visible here (dashed vertical line in Fig.~\ref{fig:adist}). 
No star with a period shorter than 0.278 days was found to have 
an $f_{68}$ frequency. 
This cut-off seems to be a peculiarity of our sample, as both in OGLE and \textit{Kepler} data 
have several stars with a period shorter than 0.278~d and yet have $f_{68}$ in their spectra \citep{Netzel2015, Netzel2015b, Moskalik2015}.

\section{Stars around the Continuous Viewing Zone}\label{sec:CVZ}

Around the ecliptic poles, all stars are continuously 
observed by \textit{TESS} until it switches hemispheres. That means we have a year's length of data.
Time series of this length provide an opportunity for other 
types of studies in addition to those discussed above.

\subsection{Stability of the main frequencies}

The period variation of RR Lyrae stars has long been a core topic of research on these stars. 
The investigation of long data series,
sometimes centuries long, has shown us that fundamental mode pulsating (RRab) stars, which do not show the Blazhko effect, are characterized by a continuous, slow period change, which is compatible with stellar evolution models. 
However, RRab stars with Blazhko effect have much 
stronger period variations and are typically irregular in nature
\citep{LeBorgne2007, Jurcsik2001, Jurcsik2012, Szeidl2011}.
The rates of these period changes are several orders of magnitude higher
than the predicted rates of stellar evolutionary theories.

The situation with RRc stars is a bit different. Besides the continuous period changes of stellar 
evolutionary origin, quasi-periodic and irregular changes
were detected for non-Blazhko stars as well. 
The latter variations are particularly common among RRc stars with longer 
main periods (see e.g. \citealt{Jurcsik2001, Jurcsik2012, Percy2013} and further references therein).

The physical background of the long and strong period changes of RR Lyrae stars 
is still unknown. For a few (quasi)periodic cases, binarity has been raised 
as a possible explanation \citep{Derekas2004, Sodor2017, Li2022}, but there are many arguments against it
(see e.g. \citealt{Skarka2018}), 
not to mention irregular changes, which certainly cannot be explained by binarity. 
Since the stellar pulsation is not a strictly repetitive phenomenon, the
accumulation of random phase variations can cause both quasi-periodic and 
irregular O$-$C (observed minus calculated) variations (see \citealt{Sterken2005} for a review). 
This random walk explanation is plausible, and it was raised a long time ago \citep{Balazs-Detre1965}.
The problem is that the magnitude of the recently verified random cycle-to-cycle  
variations in the light curves are much smaller than what could explain the observed diagrams \citep{Derekas2012,Benko2019}.
An explanation based on a similar principle is when the pulsation period fluctuates periodically 
or quasi-periodically around an average. In this case, similar (quasi-periodic or irregular) O$-$C diagrams 
could be obtained \citep{Sterne1934, Lombard1993}. This idea can explain the behaviour of 
the O-C diagrams of Blazhko RR Lyrae stars, but not that of the RRc stars that do not show the effect.

\begin{figure*}
\includegraphics[width=0.49\textwidth]{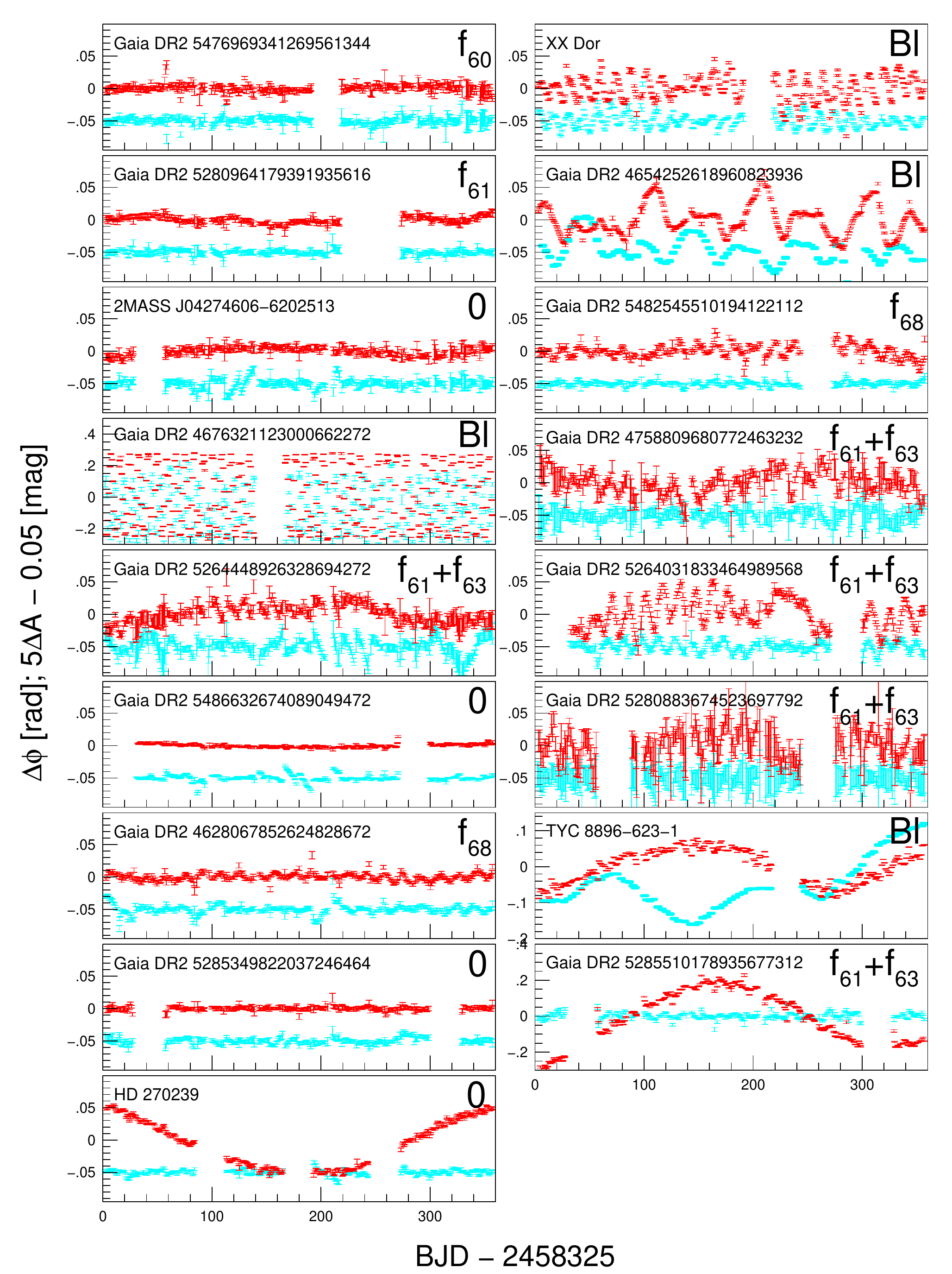}
\includegraphics[width=0.49\textwidth]{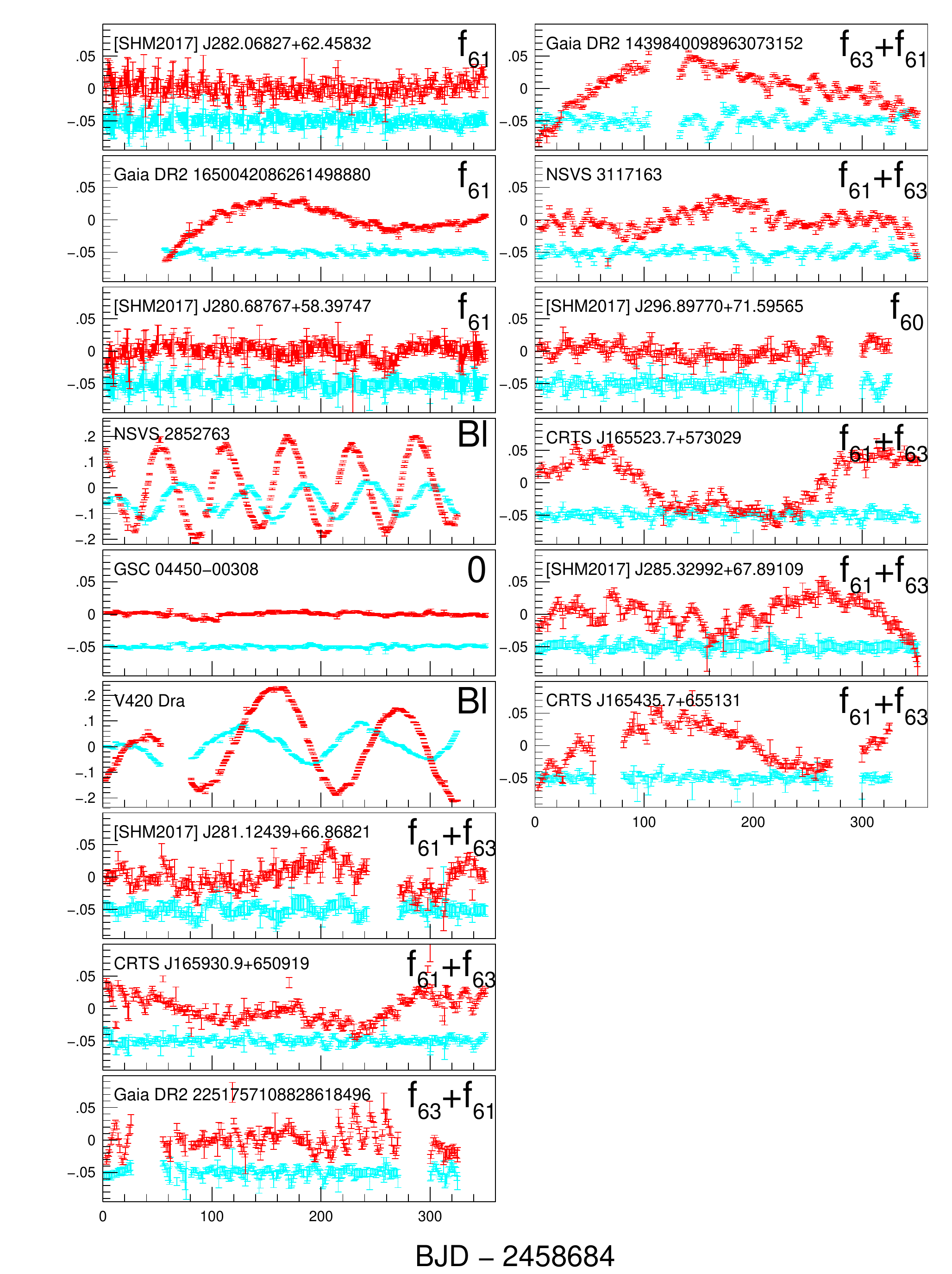}
\caption{Relative phase variation $\Delta \phi(t)$ 
of the \textit{TESS} RRc stars around the ecliptic poles. 
The light curves were collected almost continuously for a year. The two columns of panels on the left contain the stars of the southern hemisphere, while the northern stars are shown in the two columns on the right. 
The stars are arranged according to the main pulsation periods 
from top to bottom and from left to right. The time span is the same for all stars and fixed at 358 days, according to
the longest data sets. The horizontal scales are also the same (0.2 rad)
for most of the stars, except Gaia DR2 4676321123000662272, TYC\,8896-623-1, 
Gaia DR2 5285510178935677, NSVS\,2852763 and V420 Dra where larger scales are needed.
For comparison, the relative amplitude variation, $\Delta A(t)$ 
is also shown with light blue symbols in the same (0.2 mag) scale. For better visibility, $5\Delta A - 0.05$  
have been plotted.}\label{fig:phase_var}
\end{figure*}

\begin{table*}
\centering
\caption{Frequencies detected in the phase variation function spectra (northern sky, sorted by the 
main period according to Fig.~\ref{fig:phase_var})\label{tab:Fr_phase}}
\begin{tabular}{clllll}
\hline
Name  & Freq. & $S_f$ & Err. & ident. & $p_{\rm I}$\\
 & d$^{-1}$ &  & d$^{-1}$ &  & \\
\hline
[SHM2017] J282.06827+62.45832 & 0.07098 & 12.51 & 0.00081 & i? &  0.9988 \\  
 & 0.06761 & 9.64 & 0.00092 &  i & 1.0000 \\
 & 0.43823 & 5.85 & 0.00118 & &\\  
 & 0.00682 & 5.59 & 0.00121 & i & 1.0000\\
 & 0.00332 & 10.75 & 0.00087 & i & 1.0000\\
 & 0.13163 & 5.62 & 0.00120 &  & \\
 & 0.29157 & 6.02 & 0.00116 & &\\
& $\dots$ & & & \\
\hline
\end{tabular}

The entire table is available in machine-readable format as online supplement.
\end{table*}
%Ez a oc_phi.N.tex. Az additinal_3.0.zsh gyartja

Additionally, numerous RRc stars show short time-scale 
`phase jumps' (e.g. \citealt{Wils2007, Wils2008, Odell2016, Berdnikov, Benko2021}). 
This means that although the shape of the light curve is not changing, phase jumps over time must be assumed for folding
the curves properly when constructing the phase diagram. 
The unchanging light curve shape makes it unlikely that the phenomenon is caused by a long-period Blazhko effect, since 
in that case there would be a correlated change in amplitude and phase. Similar phase jump phenomena have been observed in classical Cepheids \citep{csornyei-2022}, but
have not been reported for RRab stars.

The investigation of the details of the phase jump phenomenon
was not possible in the past, since for this purpose, continuous time series are needed. 
The first opportunity to study this effect was the 4-year continuous measurements of \textit{Kepler}.
These suggest that the phase changes occur with 
a few hundred days of continuous variation and they are not sudden jumps \citep{Moskalik2015, Sodor2017}. 
In other words, it seems that phase jumps are just a manifestation of fast period changes.

During its original mission, the \textit{Kepler} space telescope observed five RRc stars
continuously for 4 years. \textit{TESS} has given us the opportunity to 
extend the study to a larger sample, as stars around the ecliptic 
pole have been measured continuously for about a year.
We selected stars from our sample that were observed by the space 
telescope in at least 10 sectors. This resulted in 32 stars, 
17 in the southern and 15 in the northern hemisphere, respectively.

\subsubsection{Phase variation functions}

A previous study on \textit{Kepler} RRab stars showed 
that the information content of the phase variation function is the same 
as that of the O$-$C curve \citep{Benko2019}. Thus, to investigate phase/period variations of RRc stars 
on an annual time-scale, we constructed their phase variation functions by a template fitting method.
To our knowledge, such method was first used for this purpose by \citet{Jurcsik2001}, 
who demonstrated its higher sensitivity compared to the classical O$-$C analysis. 
Template fitting algorithms were subsequently used as an alternative to O$-$C analysis in a number 
of publications (e.g \citealt{Derekas2012, Li2014, Benko2016, Benko2019}).

The specific algorithm we used defines a template for each star by using the 
Fourier solution of the complete data set with the main pulsation 
frequency $f_1$ and its $n$ harmonics. Then it splits 
the data into cycles and fits the template to the first
data slice containing $n_{\rm s}$ cycles. In the next step, 
the starting point of the slice to be fitted is shifted by $S$
cycles, and then this process is continued until the end of the data series.
The fit has three free parameters: the phase, amplitude and zero point of 
the light curve. We stress that we are not concerned with Fourier amplitudes and phase, 
only with global parameters. The actually used parameters were:
$n_{\rm s}=5$, $S=3$ and $n=10$, for the
number of pulsation cycles to be fitted in one step, 
the step-size in number of pulsation cycles, and 
the maximal number of harmonics to be used in the template, respectively. 
These parameters were not optimal for
TYC\,8896-623-1 and Gaia\,DR2\,4654252618960823936, for which we used $n_{\rm s}=10$.
To find the optimal values of the aforementioned global parameters in case of each light curve chunk, we used \texttt{scipy}'s \texttt{curve\_fit} method \citep{2020SciPy}. The fittings were performed using the trust region reflective algorithm \citep{trf}. The errors of the parameters were calculated from the diagonal of the estimated covariance matrix returned by \texttt{scipy}.

From each $\phi(t)$ phase change curve, we subtract a fitted linear component. 
This step both corrected for any inaccuracies in the period (which would cause a linear trend in the plots) and shifted the curves to a common mean of zero. 
We called the resulting $\Delta\phi$ the relative phase. In order to transform the amplitudes
to a common scale as well, we subtracted the average amplitude $\langle A \rangle$ from the amplitude variation functions. These became the relative amplitudes $\Delta A(t)=A(t)-\langle A\rangle$.
The panels of Fig.~\ref{fig:phase_var} show the relative phase variation $\Delta \phi$ as a function of time
for all 32 stars (red symbols) and the simultaneous change in relative amplitude $\Delta A(t)$ (light blue symbols).

Looking at the long-scale behavior of the curves, 
we can distinguish the following types: 
(1) constant phase and amplitude,
(2) annual-scale change in the phase but no significant amplitude change, 
(3) correlated periodic variations both in amplitude and phase.
In addition to all this, (4) smaller amplitude and shorter period 
phase (and amplitude) fluctuation is also seen in the curves of many stars.

Case 1 (see e.g., stars like 
Gaia\,DR2\,5486632674089049472 or
GSC\,04450-00308 in Fig~\ref{fig:phase_var})
does not require further explanation. These stars have stable period and amplitude. 

The behavior of the stars in Case 2 is similar to KIC\,4064484 and KIC\,9453114 studied by
\citet{Moskalik2015} and to KIC\,2831097 found later by \citet{Sodor2017}.
Namely, a phase change with a large amplitude and long period is seen in the stars. 
The period, if it is a periodic signal at all, is longer than the observation time span
(see e.g., HD\,270239, Gaia\,DR2\,5285510178935677312, Gaia\,DR2\,1650042086261498880,
or CRTS\,J165435.7+655131). For other stars the shape of the curve is even less pronouncedly periodic
(e.g. CRTS\,J165930.9+650919,  CRTS\,J165523.7+573029).
Since this long timescale phase variation is not associated with a similar type of amplitude change, these stars unlikely to be long-period Blazhko stars. The variations we are seeing here are probably no more than
the actual course of time the phenomenon 
which appeared in previous ground-based observations as seasonal phase jumps.

When we detect an amplitude variation correlated
with the phase variation, and the period of these variations are much longer than the main period (Case 3), 
we are faced with the classical Blazhko effect by definition. Such stars are: 
Gaia\,DR2\,4676321123000662272,
Gaia\,DR2\,4654252618960823936,
TYC 8896-623-1,
NSVS 2852763,
V420 Dra
and XX Dor.
That is six stars from the 32-element sample, or 18.7 per cent of the sub-sample. 
Consecutive Blazhko cycles are usually not identical, 
indicating the presence of more than one periods (see e.g. 
Gaia\,DR2\,4654252618960823936, 
NSVS 2852763,
V420 Dra). 
The frequent multiperiodicity of the Blazhko phenomenon 
has been demonstrated for \textit{Kepler} RRab stars by \citet{Benko2014}, 
i.e., there is no difference in this respect between the Blazhko effect of fundamental mode and overtone stars.

Let us now turn to Case 4. With such quasi-periodic phenomena, there is always the question of whether we 
are dealing with some kind of instrumental noise, especially concerning the pulsation amplitudes. 
For this reason, we emphasize that the phenomenon 
is perhaps even clearer in the phases, which are much less sensitive to instrumental problems, 
since they are primarily related to time measurements.
In Fig.~\ref{fig:phase_var}, the additional frequencies found in each star are marked in the upper 
right corner of the panels. There is a clear correlation between the strength of this type of fluctuation
and the number of exited additional modes. The  most stable stars are 
those which have no extra frequencies (marked by `0' in Fig.~\ref{fig:phase_var}), then followed by those with one mode ($\ell=8$, $\ell=9$, or $\ell=10$). Stars with 
two simultaneously exited ($\ell=8$ and $\ell=9$) non-radial modes have generally the highest phase (and amplitude) fluctuations.
The appearance of the $f_{68}$ mode also causes some fluctuations (see: Gaia\,DR2\,4628067852624828672 and Gaia\.DR2\,5482545510194122112), 
but to a much lesser amount than that of the $f_{61}$ or $f_{63}$ modes.
One aspect of this phenomenon has already been reported in the work of \citet{Moskalik2015}, namely, 
that the amplitude variation of the $f_{61}$ frequency over time is accompanied by amplitude variation in the main frequency over time. Because of our larger sample, it became clear that these phase (and amplitude) fluctuations are caused by the appearance of additional mode frequencies, or more precisely, their variation over time.

In Fig.~\ref{fig:phase_var}, the stars of the two hemispheres are sorted separately according to their main periods. 
Apart from the Blazhko stars, there is a tendency that stars with longer periods show 
phase fluctuation more frequently than stars with shorter periods. This is more evident from the left 
two panels in Fig.~\ref{fig:phase_var} (southern hemisphere stars).
This is the same phenomenon that we have already discussed in the Sec.~\ref{sec:distr_add_types}: 
the $f_{61}$ and $f_{63}$ frequencies together typically appear in stars with longer periods 
(see also Fig.~\ref{fig:adist}).

As we mentioned above, the same kind of tendency was previously published between the 
large-scale irregular changes in the O$-$Cs over decades and the main pulsation period.
This raises the hypothesis that we have already found the explanation of O$-$C behaviour of 
non-Blazhko RRc stars: the phase fluctuations of the main period caused by 
the variation in the additional frequencies are summed up in the O$-$C diagrams.
In other words, non-Blazhko RRc stars that show strong O$-$C variation are most 
likely excited by at least one, but more likely two, non-radial modes. 
And stars that do not show such O$-$C variations are likely to have no non-radial modes present.
A detailed discussion of the topic will be given in a separate paper.

\subsubsection{Frequencies in phase variation functions}

For qualitative results, all phase and amplitude variation curves were subjected 
to Fourier analysis using again the {\sc SigSpec} program package. The program continuously 
pre-whitens the Fourier spectrum with the frequency of the highest significance as long as 
it finds a significant ($S_f>5$) frequency. The frequencies found in phase
variation curves are listed in Table~\ref{tab:Fr_phase}.
After the name (column 1) we show the frequencies (col.~2)
their spectral significance (col.~3) 
and their estimated $1\sigma$ error (col.~4).
The accuracy of the frequencies obtained from a discrete Fourier analysis is not a simple task to compute. 
In a semi-empirical study, \citet{Kallinger2008} demonstrated that 
the value $\sigma_K=1/(T\sqrt{S_f})$ is a reliable upper estimate for the frequency determination error
so we used this estimate in this section. (Here $T$ is the total time span of the observation.)
These $\sigma_K$ values are given in col.~4 of Table~\ref{tab:Fr_phase}. 
In our case these values are between 0.00032 and 0.00147~d$^{-1}$.
The remark column of Table~\ref{tab:Fr_phase} contains possible identifications of the frequencies. 
Since the amplitude is sensitive to technical issues we do not discuss the Fourier spectra of the
amplitude variations in detail. We just note with an asterisk here in column 5 if the 
given frequency is significant in the spectrum of the amplitude change curve as well.

The spectrum of star Gaia\,DR2\,5285349822037246464 contains
no significant frequencies but for all other stars 2--15 significant frequencies have been found.
In total, 203 frequencies were found in the 32 stars.
Many similar frequencies appear in different stars. Some of these are certainly not of stellar but of instrumental 
origin (e.g., caused by inappropriate stitch of sectors, instrumental trends, data length etc.). 
The critical point is in which cases can two frequencies be considered identical. 
If the difference between two frequencies from different stars is less than  $1 \sigma_{\rm L}$
(where $\sigma_{\rm L}$ is the higher standard 
deviation of the two frequencies), then they should be considered identical. 
If we consider these identical frequencies (possibly from more than two stars), there is a possibility that these could be instrumental frequencies.

We selected those $f$ frequencies that could be detected in at least two stars as possible 
instrumental frequencies, and then estimated the numerical probability that they are. The Nyquist frequency $f_{\rm N}$ of the phase curves 
of each star varies between 0.666 and 0.412~d$^{-1}$
as the sampling frequency of the phase curve is inversely 
proportional to the pulsation period.
The distance of possible pairs of frequencies found for a given 
star's frequency $f$ is: $\Delta f_1=\vert f-f^{(1)}\vert$, 
$\Delta f_2=\vert f-f^{(2)}\vert,\dots,\Delta f_k=\vert f-f^{(k)}\vert$,
where $k$ is the number of frequency pairs.
For example, if $k=1$, it means that for a frequency $f$ of a given star, 
we have found a frequency $f^{(1)}$ on another star, and their distance is $\Delta f_1$.
The probability of this situation, using the formula of the geometric distribution:
\begin{equation}\label{eq:p1}
    p=\frac{\Delta f_1}{f_{\rm N}}
    \left[\frac{f_{\rm N}-\Delta f_1}{f_{\rm N}}\right]^{m-2}.
\end{equation}
Here $m$ is the number of frequencies found on all the other stars in the hemisphere.
By elementary considerations, Eq.~(\ref{eq:p1}) can be generalized to the case where $k$ 
frequency pairs can be identified:
\begin{equation}
    p=\left[\prod_{j=1}^k\frac{\Delta f_j}{f_{\rm N}^{(l)}}\right]
    \left[\frac{f_{\rm N}^{(l)}-\sum_{j=1}^k\Delta f_j}{f_{\rm N}^{(l)}}\right]^{m_l-k-1},
\end{equation}
where $l$ indexes the different stars.
The value of $p_{\rm I}=1-p$, the probability that a given frequency is instrumental in origin, is shown in column 6 of Table~\ref{tab:Fr_phase}.
If $p_{\rm I}\approx1.0$, we considered the frequency to be instrumental (sign `i' in col. 5 of Table~\ref{tab:Fr_phase}).
We identified 125 such frequencies (62 per cent).
If  $0.9999>p_{\rm I}>0.51$ then it is possible that the frequency is instrumental and an `i?' mark is shown
in the table. An additional 39 such frequencies were found.
Based on the fact that they appear in the spectra of several stars, 164 (81 per cent) of the total 203 frequencies found could be instrumental in origin.

\begin{figure}
\includegraphics[width=0.45\textwidth]{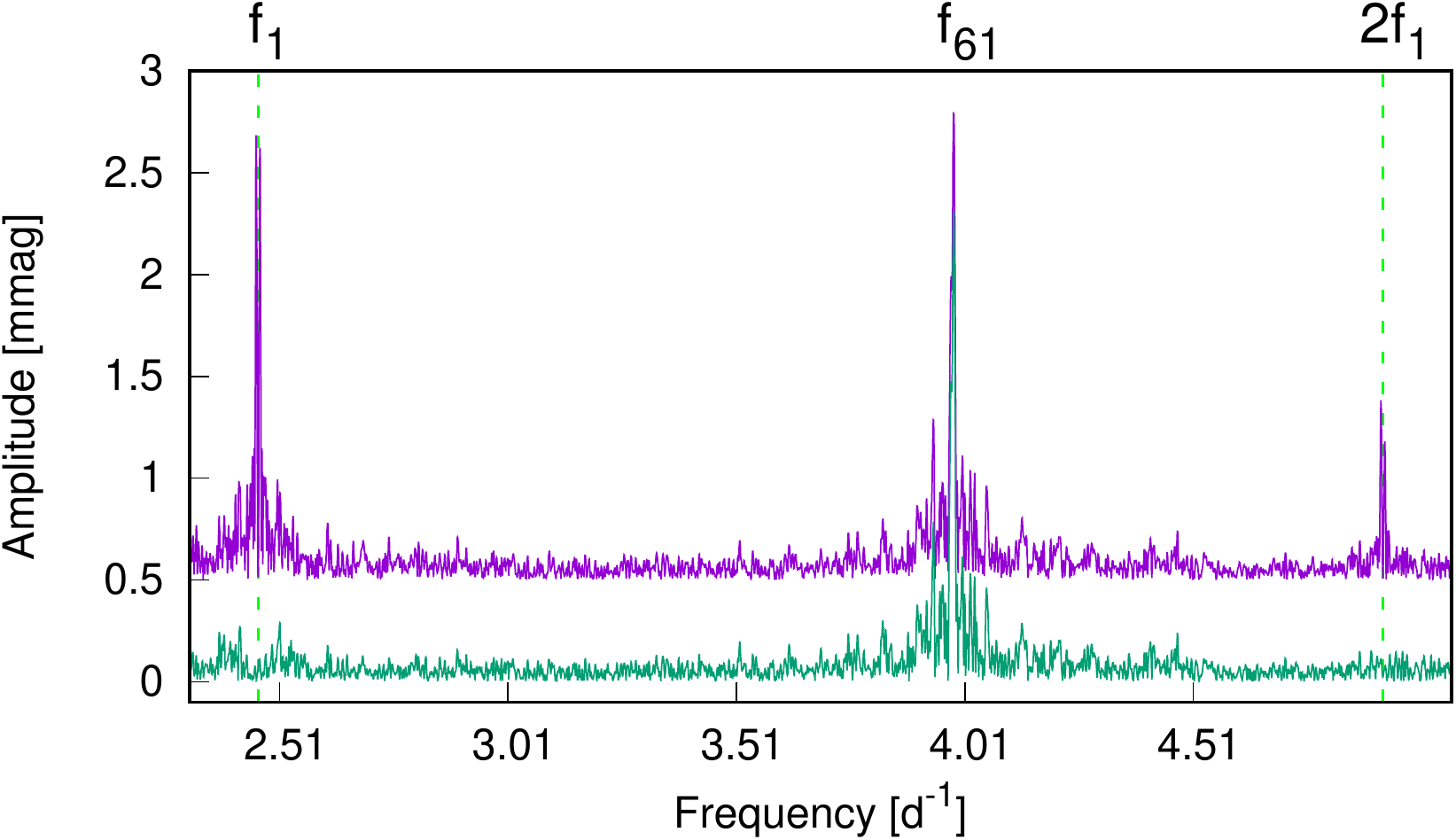}
\caption{Pre-whitened spectra of CRTS\,J165435.7+655131 between $f_1$ and $2f_1$.
The upper (magenta) line is the spectrum of the original light curve (shifted by 0.5 mmag for better visibility), 
while the bottom (green) line shows the spectrum
of the light curve from which we have subtracted the time variation of the main period.}
\label{fig:corrected}
\end{figure}

\begin{figure}
\includegraphics[width=0.5\textwidth]{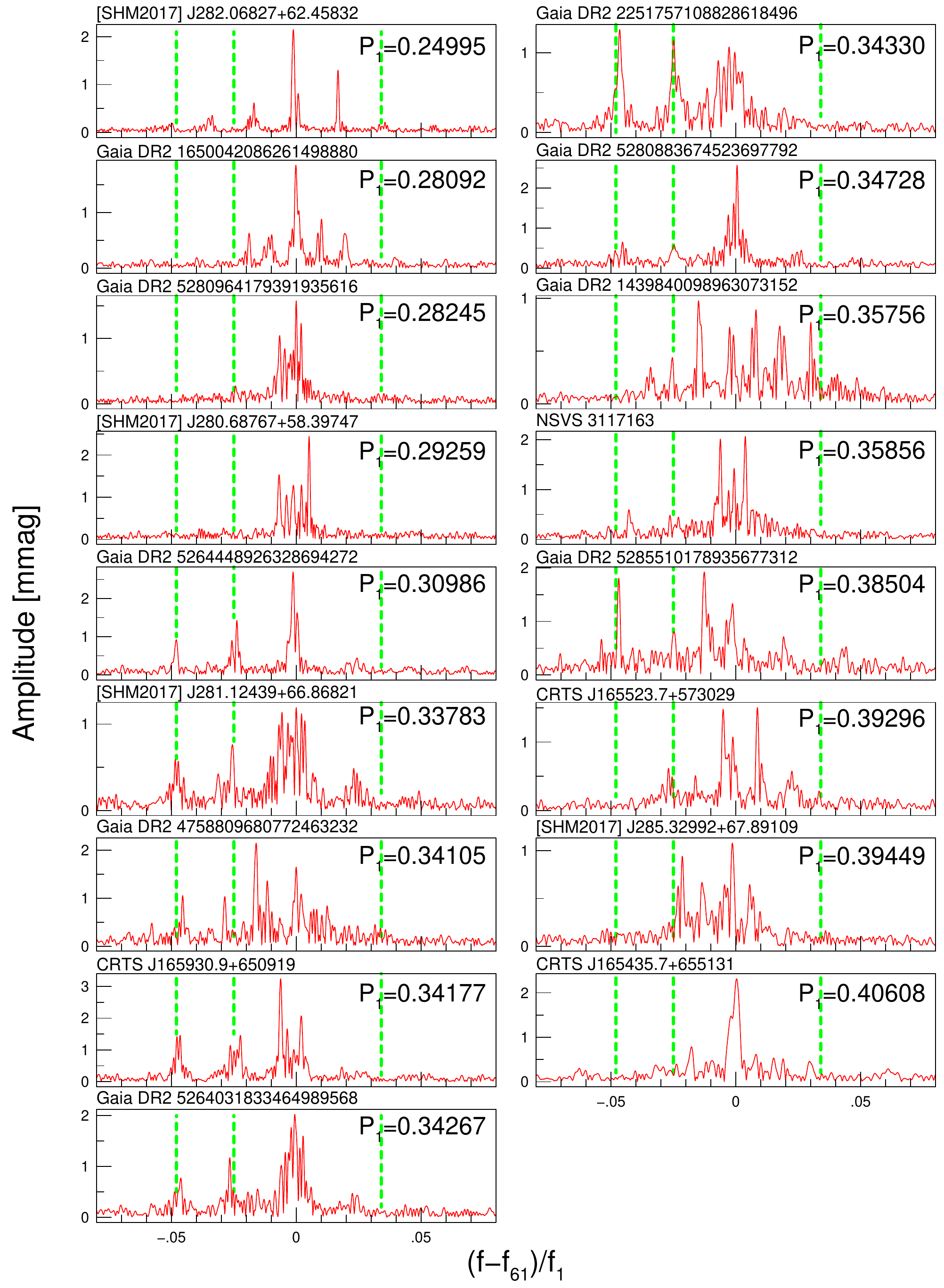}
\caption{Fine structure of the pre-whitened spectra around the additional frequency $f_{61}$
after the main frequency variations have been removed. 
For comparison, the horizontal scale is normalized by the main frequency.
The vertical green dashed lines from left to right show the possible average position 
of the frequencies $f_{63}$, $f_{62}$ and $f_{60}$. 
\label{fig:split}}
\end{figure}

\begin{figure}
\includegraphics[width=0.47\textwidth]{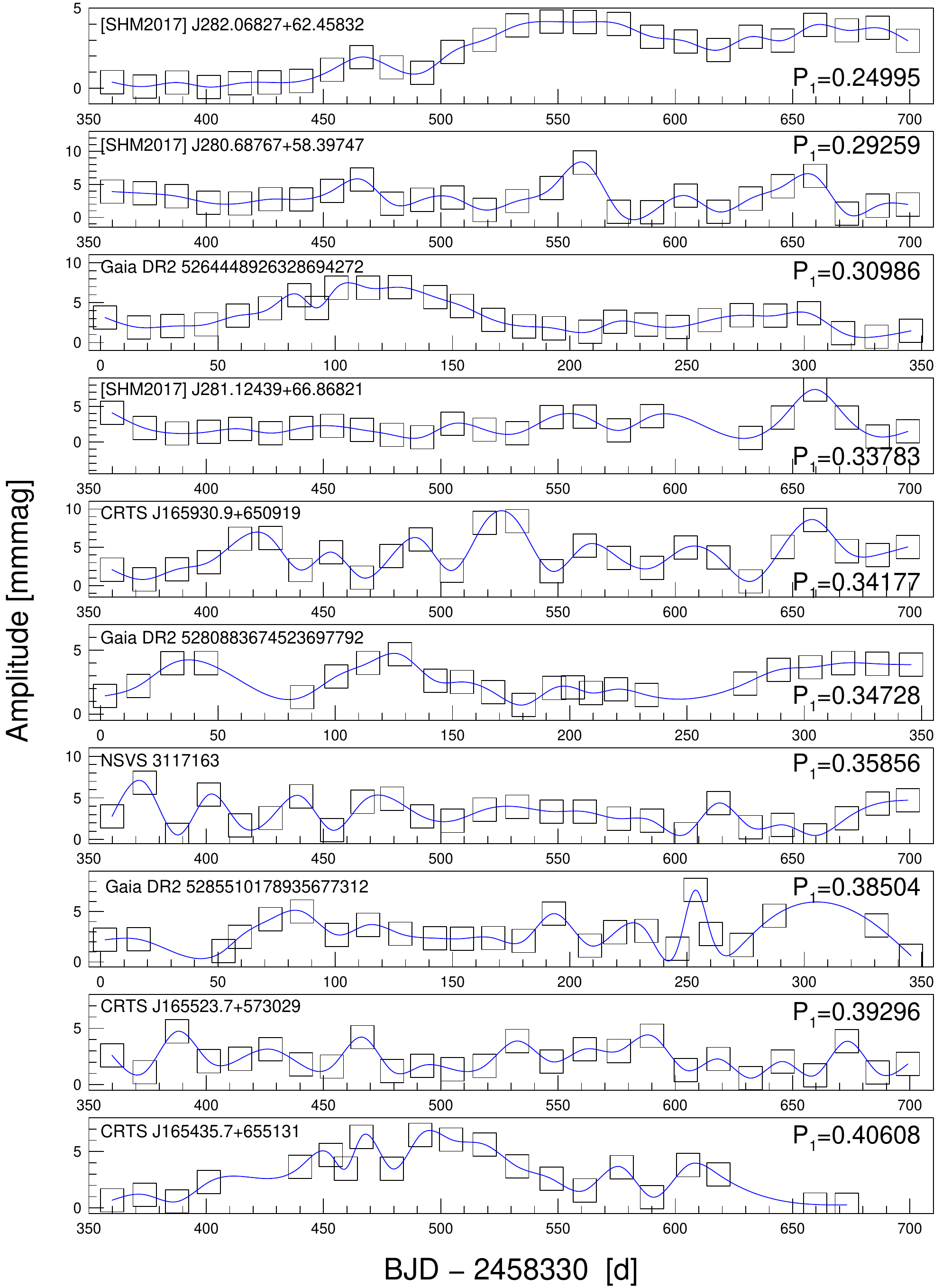}
\caption{ 
Time dependence of the amplitudes of the $f_{61}$ frequencies. 
The main period increases from top to bottom. The empty rectangles symbolize the amplitude 
and its error calculated on a given time segment. 
The continuous blue lines are spline interpolations, which are shown to make it easier to 
follow the changes.
}
\label{fig:f61_var}
\end{figure}

Of course, this is an upper limit estimate, since intrinsic 
frequencies can also coincide with each other or with instrumental frequencies.
For instance, the long time-scale phase variation (Case 1 and Case 2) 
can formally be described by a few low frequencies and such 
frequencies can really be detected in all these stars.
However, these frequencies are primarily determined by the 
length of the data series, and as such are considered instrumental frequencies.

In stars showing the Blazhko effect (Case 3), the Blazhko frequency 
(and sometimes its harmonics) can be detected.
The multi-periodic cases were also reflected in the spectra, in which more than one significant Blazhko frequency could be identified.

The low amplitude and higher frequency fluctuation in the curves (Case 4) 
yield the most significant frequencies. The characteristic time 
of these fluctuations is a few days which often makes it difficult 
to separate them from variations of technical origin
such as satellite orbital period (13.7~d) or sector length ($\sim27.4$~d). Typically, there are several similar frequencies in the phase variation curve of a single star.
The appearance of close frequencies explains the beating phenomenon observed in several stars
(see i.e. Gaia\,DR2\,5264031833464989568, Gaia DR2 251757108828618496).
This strengthens the claim that these variations are quasi-periodic.

\subsection{Time variation of additional modes}\label{sec:t_add}

As we have seen above, the amplitude and phase of the main period vary in time in most stars.
This means that before we study the time dependence of the additional frequencies, 
we need to separate this variation, otherwise we would see a mixture of two effects.
The time dependence of the main pulsation appears in the Fourier spectra as peaks around the main 
frequency and its harmonics, which remain in the residual spectrum after a standard pre-whitening step.
\citet{Moskalik2015} solved the problem by introducing time-dependent Fourier analysis. Here we 
have chosen a less sophisticated, but still functional method: we have subtracted 
the amplitude and phase variation curves of Fig.~\ref{fig:phase_var} from the light curves, 
and then analysed the residual light curves.
In practice, the amplitude and the phase modulation and the zero point curves are binned to 30 points 
and then interpolated to the original time steps. This makes it easier to track the complex 
long-period variations than, for example, a polynomial fitting.
The efficiency of our method is shown in Fig.~\ref{fig:corrected}, where we present the 
pre-whitened spectrum of the original and corrected light curves of a typical star.
The residuals around $kf_1$ almost completely disappear from the corrected data series.

\begin{figure}
\includegraphics[width=0.22\textwidth]{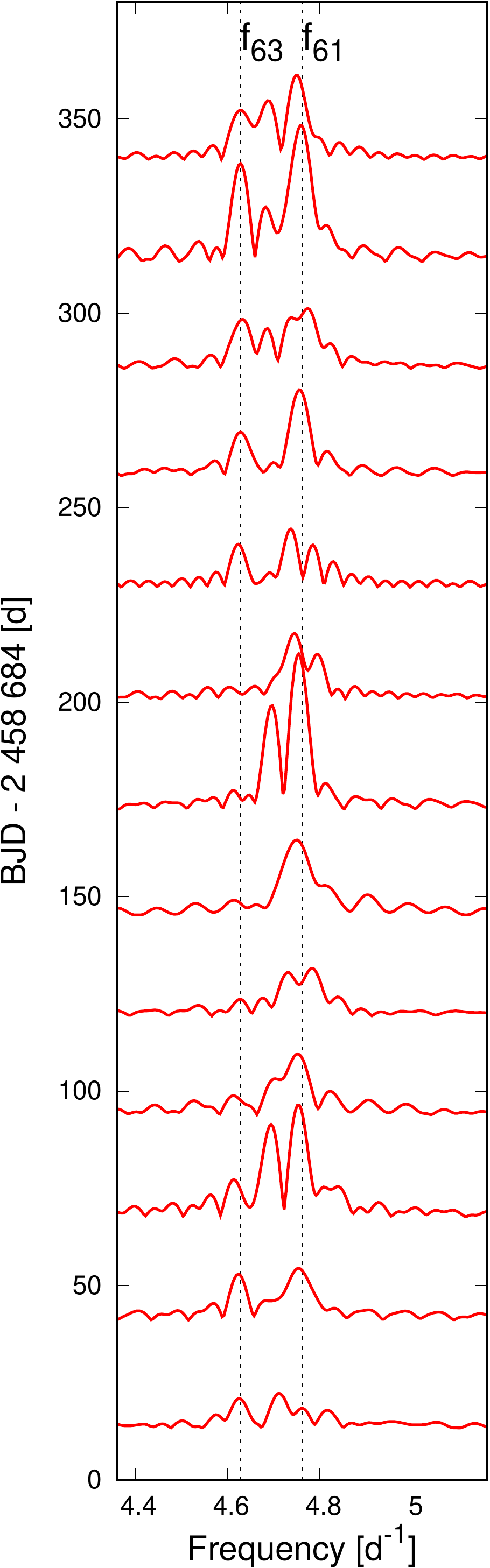}
\includegraphics[width=0.22\textwidth]{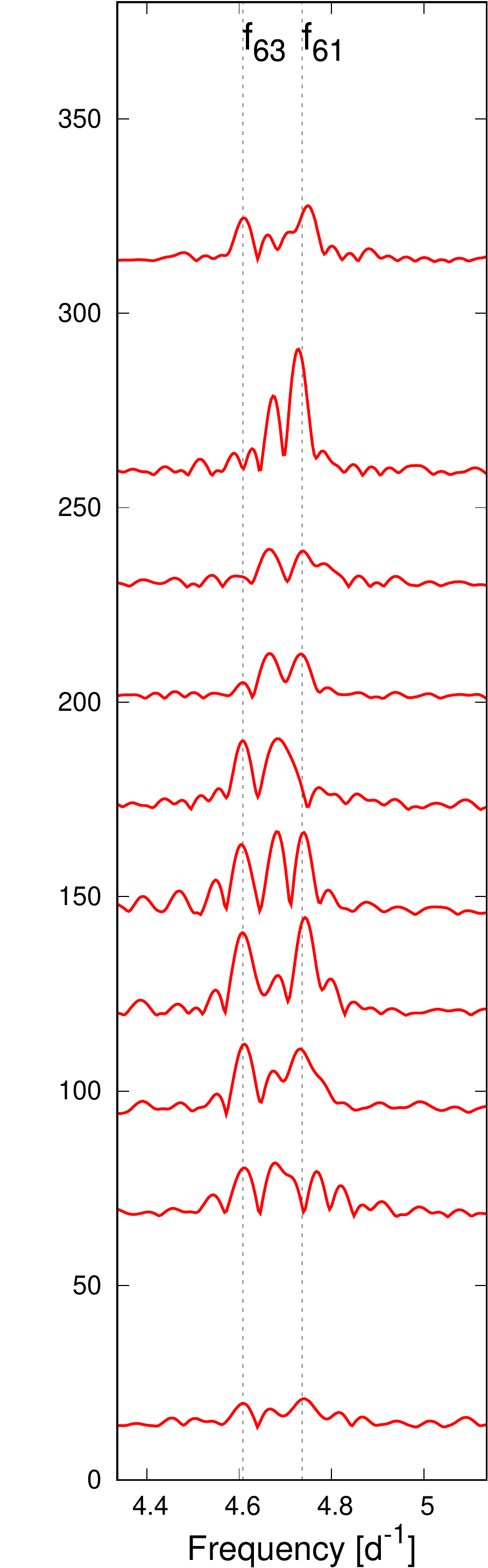}
\caption{Time variation of the spectrum around $f_{61}$ and $f_{63}$ frequencies.
left: CRTS\,J165930.9+650919, right: Gaia\,DR2\,2251757108828618496. The dashed vertical lines show the
position of the average frequency of $f_{61}$ and $f_{63}$, respectively.
\label{fig:f61var}}
\end{figure}

From the corrected data series, we selected those that contain the $f_{61}$ frequency but do 
not show the Blazhko effect. 
Fig.~\ref{fig:split} shows the pre-whitened spectra of these stars around the frequency $f_{61}$. 
The stars are sorted by their main period in ascending order from top to bottom.
Starting with the shortest period star, we see that in the first 3-4 cases, 
as the main period $P_1$ increases, the characteristic time scale of the
quasi-periodic variation of the $f_{61}$ frequency decreases. 
This is the same effect that \citet{Moskalik2015} found for \textit{Kepler} RRc stars.
This is shown by the fact that 
the `side peaks' caused by this variation
are getting wider and wider from the central peak of $f_{61}$ 
similarly to how the true side peaks of a strictly periodic modulation behave.
The positions of the frequencies associated with other modes are marked with vertical green lines to separate them from the side peaks. 
(In the first two panels, the symmetrical side peaks further away from the main peak 
are the instrumental peaks associated with the TESS orbital period 13.7~d.)
If we continue to examine the spectra with increasing main period,
we find that (i) the $f_{63}$ frequency appears from Gaia DR2 5264448926328694272 (leftmost peaks), 
and (ii) the distance to the side peaks around the $f_{61}$ peak does not change continuously. 
Point (i) is in agreement with the picture obtained in 
Sec.~\ref{sec:distr_add_types}: RRc stars with longer periods typically excite both 
$\ell=8$ and $\ell=9$ non-radial modes.
While point (ii) indicates that there is no direct relationship between the 
variation period of the $f_{61}$ frequency and the main period. 

Let us split the corrected and pre-whitened residual light curves according to the gaps in them, 
and thus calculate amplitudes on data series of roughly one orbit length with 
{\sc Period04}. The results for ten selected stars are shown in Fig.~\ref{fig:f61_var}.
The width of the rectangles in the figure represents the time interval used 
and the height the amplitude error obtained. 
Note that the construction of such diagrams is not always straightforward, 
as already shown by \citet{Szabo2014} (see their figure~11.), 
the $f_{61}$ frequencies are often split and it is not always obvious which one is the 
main peak. The main period of the stars in the figure increases from top to bottom. 
There does not appear to be any clear relationship between the main period 
and the characteristic time of the amplitude changes.
 Fig.~\ref{fig:f61_var} is an alternative representation of what is found above: 
the trend reported by \citet{Moskalik2015} is rather a small sample effect.

The $f_{61}$ and $f_{63}$ frequencies appearing in the same star seem to vary differently 
in time because the structures of the side peaks around them are different. 
This is particularly evident for stars e.g., CRTS\,J165930.9+650919 or Gaia\,DR2\,2251757108828618496.
In Fig.~\ref{fig:f61var} we show the time evolution of the Fourier spectra of these two stars 
around the frequencies $f_{61}$ and $f_{63}$. 
The spectra were calculated for each sector separately
and plotted by shifting the spectra to the midpoint of the observation time of each sector.
It is clearly visible how the amplitudes and frequencies (phases) of the $f_{61}$ and $f_{63}$ components vary over time.
In both cases shown here, but also in general, the frequency variation of $f_{63}$ is much smaller 
than that of $f_{61}$. The amplitude of the $f_{61}$ frequencies vary in shorter time
scales than the $f_{63}$ frequencies which could already be seen from the side-peak
structures in Fig.~\ref{fig:split}.

\begin{figure}
\begin{center}
\includegraphics[width=0.4\textwidth]{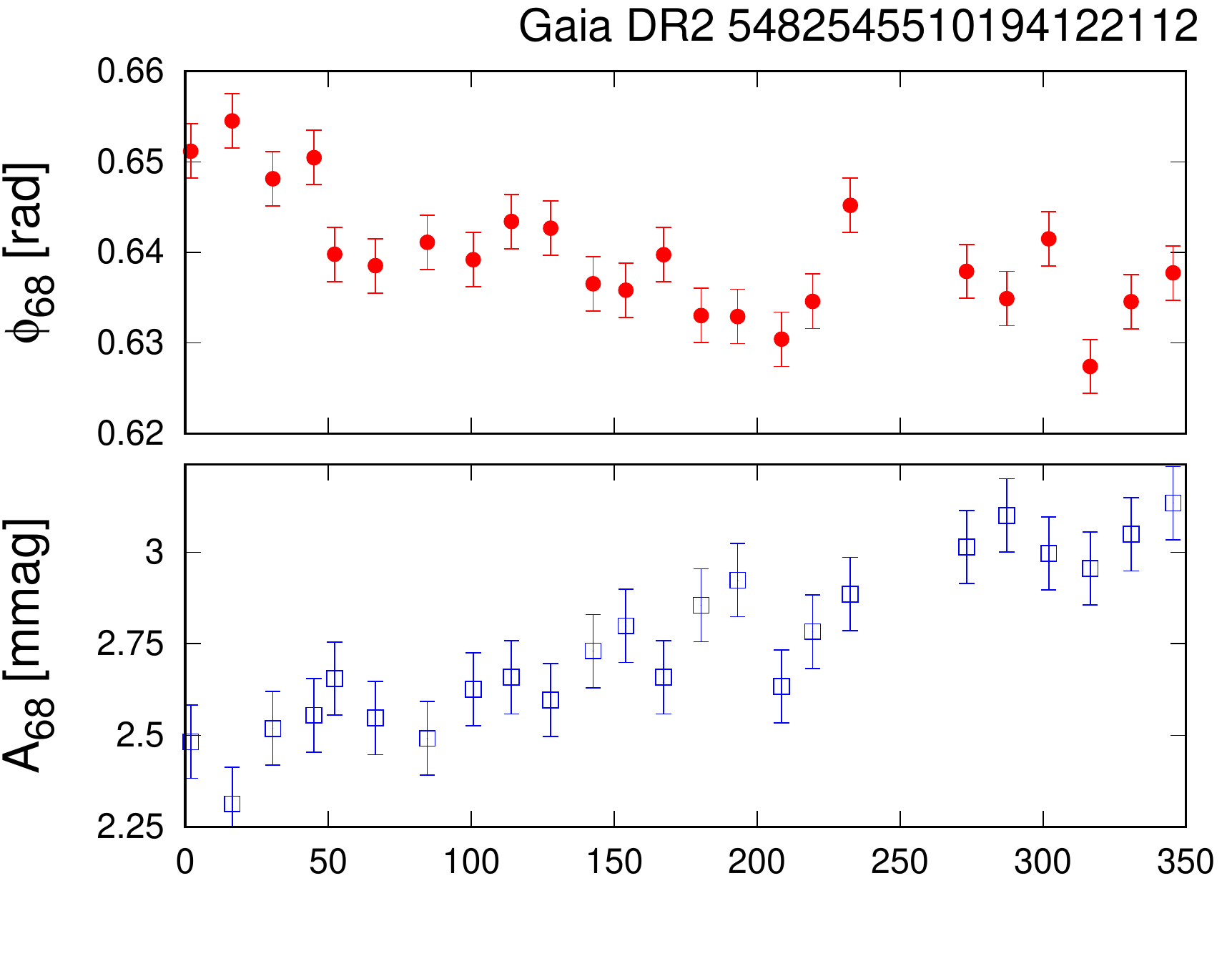}
\includegraphics[width=0.4\textwidth]{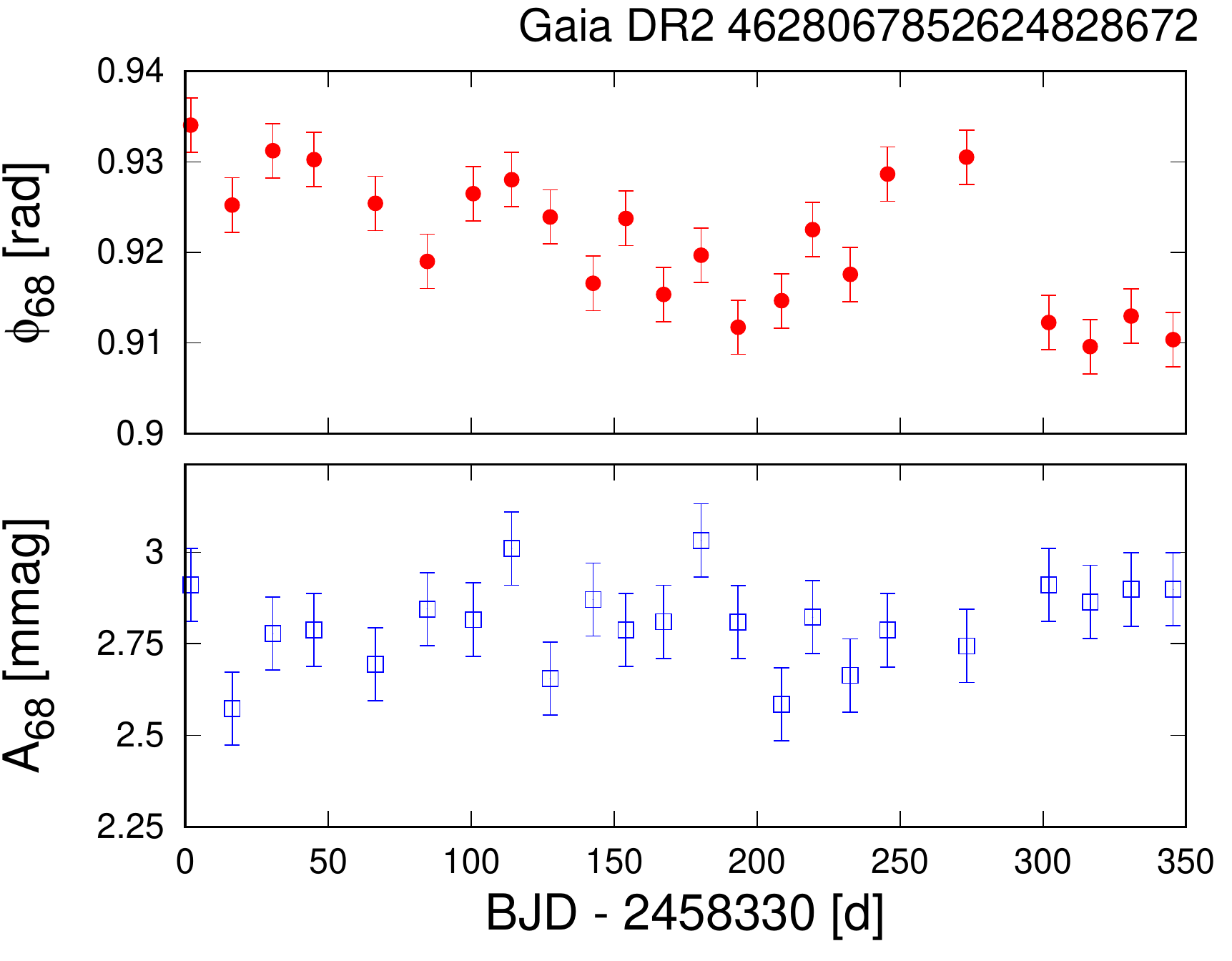}
\end{center}
\caption{The phase ($\Phi_{68}$: filled red circles) and amplitude 
($A_{68}$: empty blue rectangles) variation of the $f_{68}$ 
frequency in the two non-Blazhko stars 
(Gaia DR2 5482545510194122112 and Gaia DR2 4628067852624828672) observed in the CVZ.}
\label{fig:f_68var}
\end{figure}
The time dependence of the $f_{68}$ frequency has not been investigated in detail so far, 
although \citet{Netzel2019} noted that the frequency is coherent: ``Amplitudes
and phases of these signals do not vary in time."
We have two non-Blazhko stars in the CVZ, which show the $f_{68}$ frequency
(Gaia\,DR2\,5482545510194122112 and Gaia\,DR2\,4628067852624828672).
In a similar way to the construction of Fig~\ref{fig:f61_var},
we split the light curves according to the gaps in them, 
and then calculate both amplitudes and phases with {\sc Period04}.
The results are shown in Fig.~\ref{fig:f_68var}.
The amplitude and phase of the $f_{68}$ frequency in Gaia\,DR2\,5482545510194122112
varies in a  moderately anti-correlated manner 
(the Pearson coefficient is $-0.623$).
The variation is small in both cases: $\sim0.02$~rad in phase
and $\sim 1$~mmag in amplitude. 
For Gaia\,DR2\,4628067852624828672, the phase change is again on the scale of 0.02~rad, 
but the amplitude is constant within $3\sigma$ error.
Based on these two stars, we suspect the $f_{68}$ frequencies may also vary but in that 
case the variations are at least one order of magnitude lower in amplitude, 
and at least two orders of magnitude lower in phase than that of the $f_{61}$ frequencies, respectively.
For a more definite statement, a larger sample (e.g., several seasons of \textit{TESS} observations) 
needs to be investigated.

\begin{figure*}
\includegraphics[width=0.35\textwidth, angle=270]{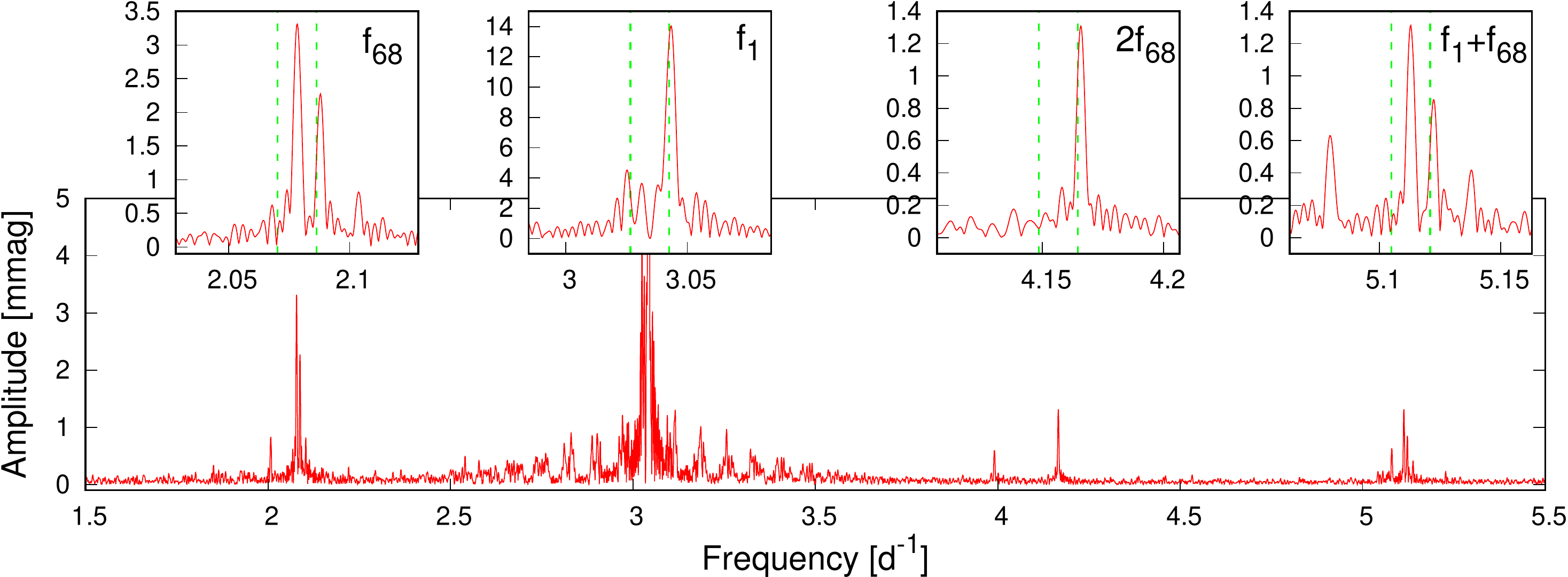}
\caption{Extract of the pre-whitened spectrum of the Blazhko star V420 Dra.
In the zoomed in panels, it can be clearly seen that the Blazhko effect is also affecting the $f_{68}$ frequency. 
The dashed vertical lines show the positions of the first-order Blazhko side peaks.
\label{fig:2f68}}
\end{figure*}

If we look at the spectra of Blazhko stars that contain the $f_{68}$ frequency, 
we see that Blazhko side peaks also appear around the $f_{68}$ frequency (see Fig.~\ref{fig:2f68}). 
For V420\,Dra, the $2f_{68}$ harmonic is not even significant, 
only the right Blazhko side peak ($2f_{68}+f_B$) is visible in the spectrum 
(the situation is similar for NSV\,2852763).
%The influence of the Blazhko effect on the $f_{68}$ frequencies further strengthens 
%that these are in fact frequencies that belong to RRc stars.

\section{Summary}

We analysed the space photometric observations of a Galactic field 
RRc sample observed by the \textit{TESS} satellite.
Our most important findings related to the main pulsation of stars are, as follows:

(i) We found an incidence rate of 10.7 per cent for the Blazhko effect.
This is consistent with the results of previous works on different subsystems of the Galaxy
(e.g., globular clusters, Bulge).
The Blazhko effect was found to be multi-periodic in many cases 
which has already been reported in the work of \citet{Netzel2018} from ground-based data.
The same was found for the \textit{Kepler} RRab stars, 
i.e., the Blazhko effect behaves in the same way in RRc stars.  

(ii) Investigation of the larger sample of \textit{TESS} stars in the continuous viewing zone around the
ecliptic poles confirmed \textit{Kepler}'s discovery that 
the so-called `phase jumps', reported in several RRc stars 
is nothing more than a annual-scale continuous phase change and it is not a sudden jump. 
The physical origin of this type of phase change is unknown.

(iii) For many stars, the main frequency shows quasi-periodic phase fluctuations. 
This fluctuation is clearly related to the additional frequencies present in the star. 
It is the strongest when two additional modes are excited in parallel in the star.
The summation of this fluctuation over time may explain the strong, quasi-periodic or
irregular O$-$C variations published in many non-Blazhko RRc stars.

Previous space observations, and later ground-based surveys have identified several 
low-amplitude additional frequencies in the Fourier spectra of RRc stars.
In our sample, we found all of these.

(iv) The incidence rate of stars with additional frequencies was found to be 82.8 per cent. 
Stars that do not show additional frequencies are more common among shorter-period RRc stars, 
while they are rare among longer-period stars. 
This is true not only for the frequencies associated with the $\ell=8$ and 
$\ell=9$ non-radial modes, but also for the $f_{68}$ frequency, which is of unknown origin. 
For the $\ell=8$ and $\ell=9$ non-radial modes, this behaviour is consistent with 
theoretical calculations.
The situation is less clear regarding our finding that stars with shorter periods and larger 
amplitudes typically contain only $\ell=9$ mode frequencies, while stars with longer 
periods and smaller amplitudes typically show both $\ell=8$ and $\ell=9$ mode frequencies. 

(v) We have identified a new group of additional frequencies with an average
period ratio of 0.602. 
We show that these frequencies, in at least the case of stars with shorter periods, can be explained 
by the appearance of $\ell=10$ non-radial modes.

(vi) The amplitude distribution of the additional frequencies is approximately exponential, 
but there are significantly fewer stars with the smallest amplitudes and significantly 
more with the largest amplitudes than would be expected from an exponential distribution. 
The galactic distribution of stars with and without additional frequencies was found to 
be identical.

(vii) 
Our work confirms that the amplitude and phase of the $f_{61}$, $f_{63}$ frequencies vary 
in time, as shown in previous publications.
No clear relationship was found between the period of these variations and the main periods.
The time dependence of the different frequencies appearing on the same star seems to be different. 
 The variations of the $f_{68}$ frequencies are at least an order of magnitude smaller than those of the $f_{61}$ frequencies.  

\section*{Acknowledgements}
This paper includes data collected by the \textit{TESS}  mission. Funding  for the \textit{TESS}  mission is provided by the NASA Science Mission Directorate. 
This research has made use of the SIMBAD database, operated at CDS, Strasbourg, France, and NASA's Astrophysics Data System (ADS).
The research was partially supported by the  LP2018-7 Lend\"ulet grant  of  the  Hungarian  Academy  of  Sciences, the  `SeismoLab' KKP-137523 \'Elvonal and NN-129075 grants of the Hungarian Research, Development  and  Innovation  Office  (NKFIH). 
HN has been supported by the \'UNKP-22-4 New National Excellence Program of the Ministry for Culture and Innovation from the source of the National Research, Development and Innovation Fund.
JMB thanks his wife, Mrs Ildik\'o Benk\H{o}, for her valuable help.
We also thank the anonymous referee for her/his careful reading and suggestions,
which have significantly improved the manuscript.

\section*{Data Availability}

The full versions of the data tables can be found in the electronic 
supplement as tab-1.txt, tab-2.txt, tab-3.txt and tab-6.txt. 
The raw FFI images taken by \textit{TESS} satellite were accessed from Mikulski Archive for
Space Telescopes (\url{https://mast.stsci.edu/portal/Mashup/Clients/Mast/Portal.html}) while the light curves  underlying this article are available via our web site (\url{https://www.konkoly.hu/KIK/}). 
All other data generated in this research will be 
shared on reasonable request to the corresponding author.



\bibliographystyle{mnras}
\bibliography{Tess_RRc_mnras_final} % if your bibtex file is called example.bib

%\begin{thebibliography}{99}
%\end{thebibliography}

%%%%%%%%%%%%%%%%%%%%%%%%%%%%%%%%%%%%%%%%%%%%%%%%%%

%%%%%%%%%%%%%%%%% APPENDICES %%%%%%%%%%%%%%%%%%%%%

%\appendix



%%%%%%%%%%%%%%%%%%%%%%%%%%%%%%%%%%%%%%%%%%%%%%%%%%

% Don't change these lines
\bsp	% typesetting comment
\label{lastpage}
\end{document}